**Assembly Theory and its Relationship with Computational Complexity**

Christopher P. Kempes,[1*] Michael Lachmann[1*], Andrew Iannaccone, G. Matthew Fricke[2], M. Redwan Chowdhury[3], Sara I. Walker[1,3,4*], Leroy Cronin[1,5*]

[1] The Santa Fe Institute, Santa Fe, New Mexico, USA

[2] Computer Science, University of New Mexico, Albuquerque, NM

[3] BEYOND Center for Fundamental Concepts in Science, Arizona State University, Tempe, AZ, USA

[4] School of Earth and Space Exploration, Arizona State University, Tempe, AZ, USA

[5] School of Chemistry, University of Glasgow, Glasgow, G12 8QQ, UK.

*Corresponding authors' emails: ckempes@santafe.edu; sara.i.walker@asu.edu; mlachmann@gmail.com; lee.cronin@glasgow.ac.uk

**Abstract**: Assembly theory (AT) quantifies selection using the assembly equation and identifies complex objects that occur in abundance based on two measurements, assembly index and copy number, where the assembly index is the minimum number of joining operations necessary to construct an object from basic parts, and the copy number is how many instances of the given object(s) are observed. Together these define a quantity, called *Assembly*, which captures the amount of causation required to produce objects in abundance in an observed sample. This contrasts with the random generation of objects. Herein we describe how AT's focus on selection as the mechanism for generating complexity offers a distinct approach, and answers different questions, than computational complexity theory with its focus on minimum descriptions via compressibility. To explore formal differences between the two approaches, we show several simple and explicit mathematical examples demonstrating that the assembly index, itself only one piece of the theoretical framework of AT, is formally not equivalent to other commonly used complexity measures from computer science and information theory including Shannon entropy, Huffman encoding, and Lempel-Ziv-Welch compression. We also include proofs that assembly index is not in the same computational complexity class as these compression algorithms and discuss fundamental differences in the ontological basis of AT, and assembly index as a physical observable, which distinguish it from theoretical approaches to formalizing life that are unmoored from measurement.

## 1. Introduction

Understanding evolution in the most general terms, including how the process of inorganic selection manifested before the genetic code, is of primary interest for uncovering the origin of life and the prospects of other life in the universe [1–7]. A major challenge is that current theories of evolution begin from some intermediate, pre-existing biological system and as such, can only describe the evolutionary dynamics within that system. Assembly theory (AT) was developed to provide a more general framework for understanding evolution and selection allowing detection of the signatures of any evolutionary process[8], including probing stages of evolution on Earth predating genomes. By combining a formal theoretical[8–10] and experimental[11,12] approach to determine whether or not observed objects are the product of evolution, AT is opening new avenues of research for life detection[13–16], the design of synthetic life[17,18], the design of experiments to probe the *de novo* origin of life in the laboratory[19], and opportunities to look deeper into the history of life on Earth (e.g., predating the evolution of the genome)[20]. In its formalization AT includes a new, empirically measurable form of complexity, *Assembly*, that adds unique features to the field of complexity metrics and opens avenues for how complexity can be measured as a physical quantity.

As with any theory that attempts to capture broad regularities in the physical world, there is necessarily a large amount of abstraction in going from measurement to theory in the development of AT[21]. This has led to some confusion about whether what the theory proposes is genuinely new[22–25] or is already captured by other theoretical metrics of complexity developed over many decades[26,27]. Some researchers have claimed that AT is identical to several other complexity measures[22–24], but do so by pointing to mathematical formalizations that are not formally equivalent to one another, and, as we show here, are also not formally equivalent to AT. In this work we hope to clarify how AT differs formally and in practice from currently available complexity measures. Providing clarity to the discussion requires demonstrating formal uniqueness and understanding the connection between different theoretical concepts of complexity. This includes how different metrics are quantified. In the case of assembly index (AI), a critical component of AT, the metric has been demonstrated to be measurable as a physical observable in molecules, whereas other metrics have not. This requires being explicit about differences in the ontological basis of computational complexity and AT, because the latter is a an empirically grounded theory of complexity, not a purely theoretical one.

The Theory of Computation was not explicitly developed to solve the problem of the origin of life[28], nor was it even devised as a formal approach to life or to physical systems. Nonetheless, the connection between computation and living processes has been noted by many researchers over the years [29–32]. The key ideas of theoretical computer science were laid out in the seminal work of Turing[33] and Church[34]. Turing's mathematical device was simple - consisting of a look-up table, state-machine, and an infinite tape on which to perform computational operations - but with it he was able to prove foundational insights into the nature of computation. For example, he proved it was possible to build a machine that could compute any computable function, now known as a *universal Turing machine (UTM)*. Early work by von Neumann attempted to analogize the concept of universal computation to that of universal construction[35]. Inspired by Turing's ideas, von Neumann's universal constructor provides a thought experiment of an abstract machine that can construct any buildable physical object (rather than compute any computable function as a universal Turing machine does). This difference is important because one theory, that of computation, deals with the inputs and outputs of programs, how to describe these, the computational resources (time and space) required, and the complexity of algorithms for producing specific outputs. The other, that of construction, was intended to deal with objects constructed by physical processes, not data and algorithms. That is, in the latter the constraints of physical law are necessarily foundational[36]. However, von Neumann's universal constructor was restricted to computational experiments with cellular automata[37], so this distinction has so far remained mostly abstract. Self-replicating robotics has extended this idea to the physical world, but so far require prefabricated building blocks and are far from being universal constructors[38]. Assembly theory (AT) makes the connection of these ideas to physics and metrology[39] concrete, thus providing the first step toward a testable theory of selection as a mechanism for constructing objects.

In what follows, we first introduce the foundations of AT as developed to date. We subsequently establish why AT is not the same as computational approaches to life. (1) We show with explicit examples how a key component of AT, the assembly index (AI), is not formally equivalent to other measures of complexity derived from computer science and information theory. Specifically, we show, with simple but rigorous mathematical counterexamples, why AI yields quantitatively different results when compared to compression algorithms like Huffman coding and Lempel-Ziv-Welch (LZW) compression, which have been argued to be identical to AI. (2) We provide formal proofs for calculation of the assembly index of strings, showing it is not in the same computational complexity class as these compression algorithms, definitively demonstrating that no formal equivalence is possible. (3) We also discuss metrology, and how assembly theory provides a rigorous approach to the physical measurement of complexity and life detection, which has not been shown to be emulated by computational complexity measures. We discuss how empirically, the assembly index is developed as a physical observable, whereas Huffman coding, LZW compression and Kolmogorov complexity are computed from labeled data. (4) Because it is physically relevant, it should be expected that assembly, when used to detect life and evolution, will have lower false positive and false negative rates than purely algorithmic measures. Any of these points would be sufficient to distinguish AT from computational approaches to formalizing life, but we also highlight how (5) AT is not based on AI alone and reemphasize the importance of the copy-number in AT. Approximating any feature of AT with other mathematical measures does not replace the theory or invalidate the explanations it might provide. Assembly theory is being developed into a field of study, formalizing and allowing empirical tests to provide insights into the emergence of life. It thus provides a much richer intellectual landscape than some early criticisms have suggested.

## 2. Assembly Theory: A Brief Overview

Any useful approach to quantifying complexity relevant to life detection (in the lab or on other worlds) must bridge the reality gap between theory and empirical testing and validation. That is, it must engage with metrology. This is a primary reason why AT was developed in the context of a modern chemistry lab's capabilities, with its initial development as a rigorous method for laboratory studies aiming to identify molecules as biosignatures[11]. A key conjecture of AT is that "life" is the only mechanism for producing highly complex identical objects in abundance, ranging from molecules, to cells, to memes and beyond. This proposal leads to a key, testable prediction of the theory that high AI objects with a high copy-number will not be found independently of evolutionary processes (see e.g., Kempes and Krakauer[40] for a broad discussion on abstracting the living process). While many kinds of measures of molecular complexity computed on graphs have been developed over the years[41–43], the key advance offered by AT was to propose and formalize a precise conjecture on the relationship between molecular complexity and life and allow an experimental test. This observation necessarily and intentionally grounds AT in experimental science, since the threshold of the copy-number and AI that constitute a biosignature is an essentially empirical one. AT, for the first time, provides a quantitative measure of how much selective constraints have been applied to generate an object and therefore how likely it is to be associated with a living process.

Assembly theory has the following observables: copy number, $n$, and the assembly index, $a_i$ [8]. The copy number, $n$, of an object refers to how many identical objects are observed. Here it is important to clarify what an **object** is in AT, because the definition departs from most definitions in physics and other fields, including as used in other approaches to formalize complexity (we shall return to this point later). In AT, objects have the properties of being finite, distinguishable, breakable (and therefore constructable) and they must be able to persist in multiple identical copies. With this definition of an object, the assembly index $a_i$ quantifies the minimal number of recursive, compositional joining operations necessary to construct an object from its elementary parts. The exact definition of assembly index is restricted to the assembly space of physically meaningful operations, see ***Figure 1*** for an example for the molecule adenosine. As AT is developed, we expect it will find applications beyond the initial focus on covalent chemistry, including to more abstracted spaces like gene / protein sequences, language, and many types of technology.

For molecules, both the copy number, $n$ (abundance of molecular species) and $a$ (minimal number of bond-making operations, allowing reuse of parts) are measurable in the lab. Assembly index $a_i$, referred to as the molecular assembly index (MA) for molecules, can be measured using mass spectrometry, NMR, and infrared techniques[12]. For an observed configuration of complex matter (e.g., a system that can be broken apart to more basic building blocks), taking account of both the number of copies of distinguishable objects and their assembly indices allows quantifying Assembly, $A$, which we conjectured captures the total amount of selection required to produce it. Formally, for an ensemble of $N$ total (non-unique) objects, the Assembly is defined in equation 1[8]:

Eq. 1 $$A = \sum_i e^{a_i} \left(\frac{n_i - 1}{N}\right)$$

where the terms of $n_i$ and $a_i$ are the copy number and assembly index, respectively, of the $i$th distinguishable object in the ensemble. The assembly equation, Eq. 1, represents a formalization of selection and evolution in a very large combinatorial space, the **assembly space**, defined as the set of recursive, physically implementable operations from which objects can be constructed (***Figure 1***).

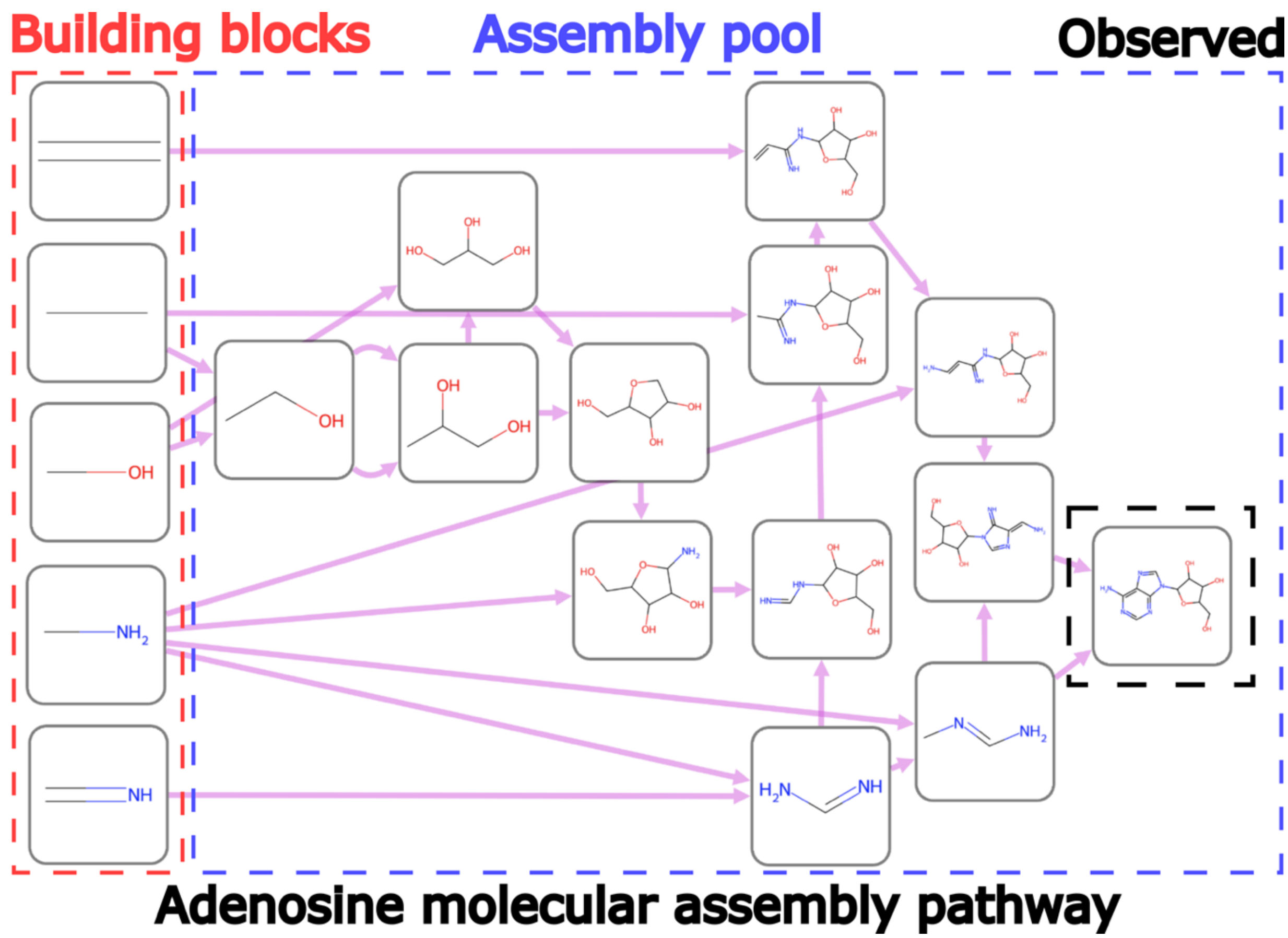

***Figure 1:*** *An illustration showing how a molecule can be cut into basic units, and recursively assembled from them to construct a minimum assembly path. The molecule's molecular assembly index is the number of steps on the path.*

If AT proves to solve the problems it aims to, the philosophy underlying the theory will lead to a reframing of many fundamental concepts including what evolution and selection are as physical mechanisms, and the objects it produces[44]. AT suggests that the combinatorial space of evolutionary objects is best understood in terms of coordinates of assembly index and copy number within the assembly space[8].

Finding objects with high assembly indices in high copy number, yields large values of $A$, implying substantial selection must have occurred[11,12]. This is because it becomes super-exponentially more difficult to produce identical copies of increasingly assembled objects with large $a_i$ by chance alone (i.e., there is a super-exponential expansion of the combinatorial space for each possible joining operation). That is, observing $n_i \gg 1$ for a complex object with sufficiently high $a_i$ is nontrivial, and in AT is conjectured to be impossible to happen abiotically. In AT, configurations of matter with large $A$ are unreasonably unlikely to occur outside of selective processes found in what we call "life". Here, we are using selection in a generalized sense: selection is the mechanism whereby a set of constructive circumstances yield formation of some subsets of objects over others. This spans simple reaction rate differences to formal modern genetics, where the strength of selection would also be larger for the latter, thus yielding higher $A$. Laboratory measurements of molecular assembly indices confirm the conjecture that an observational threshold for the evidence of selection and evolution exists. To date, experimental studies of abiotic, biological, and blinded samples indicate that molecules with $a_i > 15$ are found only in living samples[11].

## 3. Comparative Complexity and the Uniqueness of Assembly Index

The ontology underlying AT as a theory of physics is different than that of the theory of computation, see **Table 1**. In what follows, we compare AT to computational and information theoretic approaches to quantifying complexity. For some comparisons we give explicit examples, while for others we do not because the differences in ontology and computability already make clear the most salient distinctions between different approaches, and in some cases these distinctions are so apparent as to make direct comparison not meaningful. For establishing mathematical uniqueness, we take the simplest and most illustrative approach, in the form of basic counterexamples. We consider cases where the assembly index is distinct for two outputs, but those two outputs yield the same value in another complexity measure; and we show cases where the assembly index is the same for two outputs, but other complexity measures yield distinct values. Either scenario would be enough to show that the assembly index is not one-to-one with the comparison complexity measure.

| Complexity Theory | Operands | Measures | Description/Details |
|---|---|---|---|
| Algorithmic Complexity | Classes of Mathematical Problems | Computational Complexity Kolmogorov complexity; | Measures are concerned with optimal (minimal) descriptions of products of computational procedures, or the computational resources required for an algorithm that could produce the given output |
| Compression and Statistical Complexity | Data and its Representations | Huffman Code; Lempel-Ziv-Welch Compression; Information theory | Measures provide data transformations to generate reduced representations |
| Physical Complexity | States of Physical Systems | Logical Depth; Thermodynamic Depth; | Measures are concerned with the physical resource needed to compute a minimum description, or the probabilities of observed paths to produce a state |
| Assembly Theory | Physical Objects | Assembly Index | Measure is concerned with minimum construction as an intrinsic physical attribute |

***Table 1:*** *Comparative Ontology of Complexity Measures. Examples of different kinds of complexity theories and measures, describing their operands, measures and conceptual underpinnings.*

### 3.1 Assembly Index Does Not Depend on Data Representation

The most widely discussed measure of computational complexity is the Kolmogorov-Chaitin complexity[45,46], also known as algorithmic complexity, independently developed by Kolmogorov and Chaitin. Formally, the **algorithmic complexity** *K(x)* of a string *x* is defined as the length of the shortest computer program, which when run will output *x* (assuming a fixed programming language). We consider two outputs for illustrative purposes, considering the strings $x_1$ and $x_2$ where,

$x_1$ = '*353i5ab8p5fsa5jhk72i8asc47wwzlacljj9*'

and

$x_2$ = '*0123456789abcdefghijklmnopqrstuvwxyz*'.

Both strings consist of 36 characters, but the second string has a simple program description namely "print 0..35 in base 36", while the former appears random, because there is no obvious program that can produce it, other than printing the entire string, i.e., "print '353i5ab8p5fsa5jhk72i8asc47wwzlacljj9' ". Example programs that might output each of these two strings could therefore be written out as follows:

**Program 1**

print '353i5ab8p5fsa5jhk72i8asc47wwzlacljj9'

**Program 2**

print 0..35 in base 36

**Program 1** is clearly longer than **Program 2**. Defining *d(x)* as the program of minimum length that can output the string *x*, then the algorithmic complexity is defined as:

Eq 3. $$K(x) = |d(x)|$$

That is, $K(x)$ is the length of the minimum description of $x$, found by searching over all programs that output $x$ in the language $x$ is described in. The value of $K(x)$ for a given string therefore depends on the choice of description language (and for example might be different if we wrote the program in *C* versus Python). Foundational results in the theory of computation establish there can be no algorithm that determines whether an arbitrary program will produce a given output as this would involve, at least, deciding whether the program will halt. Turing's work proved this is generally uncomputable[33]. The implication is that *K(x)* is also uncomputable[45,46]. For the example above, the two strings have equivalent size, but their complexity is not the same: $K(x_1) > K(x_2)$, and $x_2$ would thus be described as more compressible. However, there is no way to guarantee $x_1$ does not have some hidden pattern in it that makes it just as compressible as $x_2$ (that is, although we have shown $K(x_1) > K(x_2)$ for the two shortest programs we were able to identify, it could still be possible there exists an algorithm K'$(x_1) \leq K(x_2)$).

In Abrahão et al.[22] and Ozelim et al.[24], the authors make an argument that since AT does not offer any innovations in approximating $K$, it is rendered redundant by pre-existing methods of lossless compression. However, this represents a fundamental misunderstanding because approximating $K$ is not, and was never, the goal of AT. It is our view that $K$ is not relevant as a *physical* measurement of complexity generated by evolutionary processes. The assembly index comes to estimate a lower bound on the chance of the random generation of the molecule from molecular inflow into the system, by formalizing minimum causation necessary to make an observed configuration. The algorithmic complexity, on the other hand, is the minimal size of the program required to generate a certain output – given a universal Turing Machine. The problem then is that the construction of the molecule requires construction of the molecule itself by the UTM (we ignore for a second the fact that a UTM cannot construct molecules), but also the construction of the UTM itself. Construction of such a reliable computer will require many more steps than the construction of the simple molecules that we are interested in when detecting life. Though it is true that very simple physical systems can exhibit universal computation when properly encoded[47], we know of no system that exhibits universal computation, can reliably control molecular reactions, and can form itself within a small number of steps. In fact, the only process that we know of anywhere in the universe that can build such a machine is the evolution of life itself. Until we can find such a process, or a reason for it to exist, the assumption that before the origin of life molecules form in large numbers only through assembly processes in the absence of computers seems well founded. Thus, the assembly index does not come to approximate $K$, instead $K$ is a bad approximation for the minimal chance of formation of molecules, a task that the assembly index was specifically designed for. As an example, let us look at a polymer that represent the first 100 digits of the number $e$ in some base, that is present in an abundance such that it can be unambiguously detected. This is a relatively simple computation, and therefore the program to compute it can be small. However, the length of this program will not represent the chance for the formation of such a molecule. Instead, it will form by assembly of the monomers plus assembly of the 100 digits, very similar to the assembly index of any other random polymer of length 100. Without a universal Turing machine, the fact that there is a small program to generate the sequence is meaningless.

### 3.2 Assembly Index is Not Reducible to Statistical Measures

Assembly index was developed as an exact, objective measure that can be probed by different experimental techniques, capturing symmetries and reuse of fragments in the assembly pathway to construct a molecule. This distinguishes it from statistical notions of complexity, which rely on computed probabilities from observed frequency of reuse of segments in a data structure. Nonetheless, the assembly index has been critiqued as both a restricted version of Huffmann coding[22,23] and described as reducible to Shannon entropy[24]. Formally, a "restricted version" or “reducible” should imply one formulation is a formal subset of another. For Huffman coding, Shannon entropy and assembly index, it is easy to explicitly show these are not equivalent and therefore that assembly index is by no formal means a 'restricted version' of Huffman coding or reducible to Shannon entropy.

Information theory provides the theoretical underpinnings for efficient data compression and coding techniques, establishing limits on to what degree data can be compressed without losing information (lossless compression), and how information can be transmitted over a noisy channel reliably. Concepts like algorithmic complexity[45,46], Huffman coding[48], arithmetic coding[49], and channel capacity[50] are all grounded in information theoretic principles. The field of information theory began with the seminal work of Shannon in 1943 using the term entropy (H) to define what is now known as **Shannon information** or the Shannon entropy[50]:

Eq. 2 $$H(X) = -\sum_{x \in X} p(x) log p(x)$$

where *p(x)* is the probability, a random variable will be in state *x*, and H(x) is referred to as Shannon entropy and represents uncertainty. “Information” is then reserved for describing a difference in uncertainty, for example, in computing how much uncertainty was reduced when a signal is received, or how much uncertainty increased due to noise in transmission. Information theoretic measures like Shannon entropy[50], mutual information[51], transfer entropy[52], and synergistic information[53] etc. are widely used to characterize the complexity, dependencies, and similarities in data across complex systems[54,55].

Shannon entropy is a powerful tool due its generality, but this also means one must exercise caution in its interpretation as it can be applied to any data labeling scheme one might wish to define[56,57]. Indeed, Shannon entropy provides a good example of how being explicit about the ontology of the abstract variables that appear in our different mathematical concepts of information is critical to understand their utility (see ***Table 1***). For example, if $X$ labels the microstate of a physical system, the mathematical form for Shannon entropy is nearly identical to the thermodynamic entropy[58] $S(X) = kTH(X)$ where here $k$ is the Boltzmann constant and $T$ the temperature of the system. This allows a physical interpretation that has utility: there are many reasons to think that many complex systems from languages to computer algorithms would target distributions of states, and the energetic costs of moving between states, as a key feature for refinement. The formal equivalence of Boltzmann and Shannon entropies has therefore led to interesting work on the connections between information theory, computation, and thermodynamics[59]. Stemming from the work of Landauer[60], much of this effort has focused on the minimal energy required to perform a computation by casting computations as entropy transformations, and on showing that storing and erasing information comes with certain energetic costs, which includes experimental verification[61]. **Huffman coding**[48] is among the most widely used lossless data compression algorithms. A Huffman code is constructed by first building a frequency table of characters and their occurrence counts, see ***Figure 2***. The compression efficacy of Huffman encoding depends on the character frequency skew, where data with highly frequent characters are compressed the most, and data with uniform distributions of characters are compressed the least.

As a simple illustrative example of formal differences between assembly index, Shannon entropy and Huffman coding, we consider the strings:

$$zbzbzca = 4$$

$$zzzbbca = 5$$

These strings are identical in length and composition (with the same number of *z*'s, *b*'s, and *c*'s), and thus have the same string entropy and Huffman coding. The two strings have different assembly indices.

The Huffman tree is constructed by creating a node for each character and repeatedly combining the two least frequent nodes/characters into a new parent node, until only one parent node, the root node, remains, see ***Figure 2***. One traverses the Huffman tree to assign bit codes to characters, assigning short codes for frequent characters and longer codes for those that are less frequent. The deeper the traversal, the longer the code. The original data is then encoded by replacing characters with their corresponding Huffman codes, writing out the concatenated bit stream. For the example above, the tree that generates the code alphabet is identical for the two strings (because they have the same character frequencies) with the following code assignments

$$z \rightarrow 1$$

$$b \rightarrow 01$$

$$c \rightarrow 00$$

which results in the encoded output strings

$$zbzbzc \rightarrow 101101100$$

$$zzzbbc \rightarrow 111010100$$

There are several observations of note here. Two strings with *different* assembly indices can produce *identical* Huffman trees and code alphabets. The two representations are not one-to-one because the Huffman trees are identical for strings that have different assembly spaces, see **Figure 2**, providing a counterexample proof that AT cannot be a subset of Huffman coding (contrary to the claims of[22]) because there can be no 1:1 map between the two applicable in every case. This establishes the assembly index is formally unique when compared to Huffman coding.

Assembly index is not a statistical measure and does not rely on the frequencies of characters, instead it aims to capture features of the recursive composition of objects as they are constructed from a base alphabet. By contrast, Huffman trees are often highly degenerate because they map the specific structure of strings onto character frequencies, which is a many-to-one mapping. In general, the assembly index is more weakly dependent on character frequencies than statistical measures, and it will yield different values in many cases where Shannon entropy or Huffman coding would not resolve differences between strings.

Some have argued that lossless compression, like that involved in generating a Huffman code, is relevant to evolution, particularly in describing genomes[62]. However, genomes are highly constrained by the evolutionary history of the ribosome, and physical constraints on protein folding[63]. As such, compression might not yield the most accurate insights into evolutionary refinement of genomic architecture, especially as it does not account for the physical and physiological constraints present in all cellular life[64–66] (or viruses[67]). Given compression-based measures are sensitive to the choice of data representation, they must be applied with care in a biological context: the labels humans use for data to describe living things may not be the same labeling scheme those organisms implement themselves, even if it were the case where an organism was indeed be performing data compression.

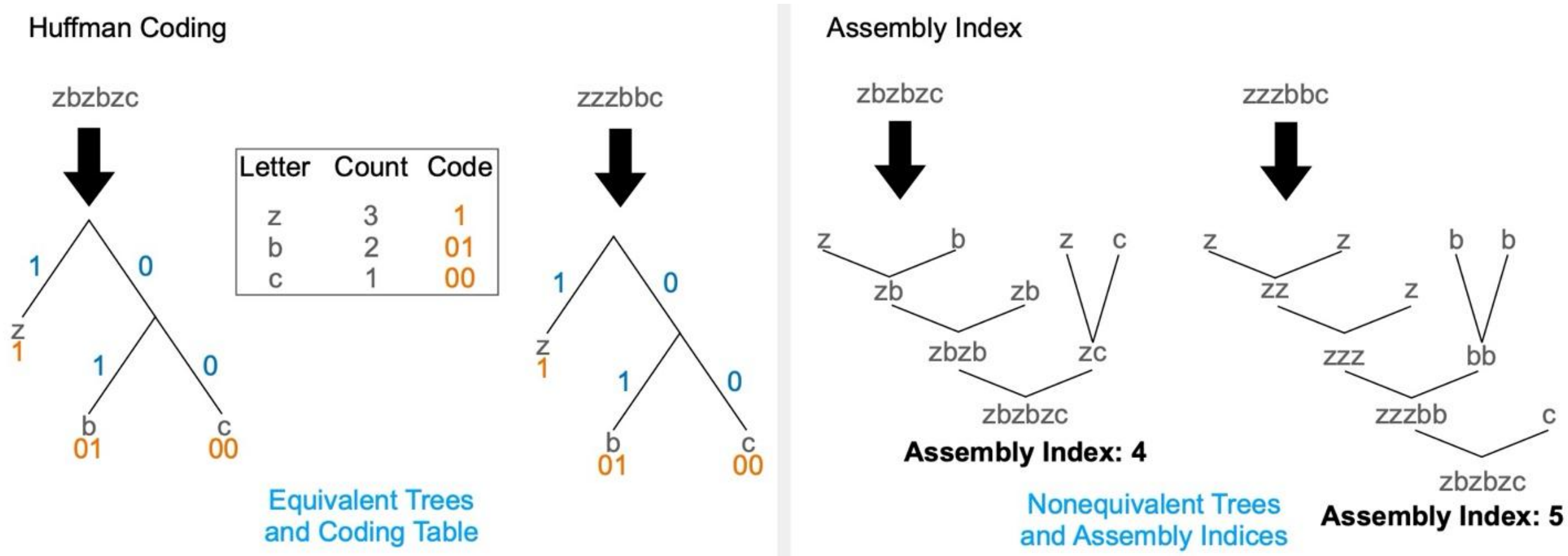

***Figure 2:*** *Two strings with different assembly index but identical Huffman trees. A single case such as this is sufficient to demonstrate that the assembly index cannot be a "restricted case" of Huffman coding, because it demonstrates the assembly indices cannot be mapped one-to-one with Huffman trees.*

By contrast, the aim of AT is different because it does not suppose a compressed description is necessary, and the theory is agnostic to assumptions that cellular life is doing any kind of computation or algorithmic compression. That is, AT does not rely on a human or computer-based labeling scheme, or any kind of statistical description of data, instead it aims to identify the underlying structure of the physical space in which selection operates (the assembly space).

### 3.3 Assembly Index is Not a Compression Algorithm

The assembly index has also been criticized by the same group of researchers as identical to other compression algorithms, including the Lempel-Ziv compression scheme[22–24]. But again, we can establish formal independence with basic worked examples for strings. We consider Lempel-Ziv-Welch (LZW) for an explicit example[68]. For LZW a string sequence is read from left to right, and a lookup table is built for new substrings. The longest matching substring is chosen, and a new substring which includes that substring and the next letter in the input is added to the lookup table along with a new index. Regular signals can be characterized by a small number of patterns and hence have low LZW complexity. Irregular signals require long dictionaries and hence have a high LZW complexity (most schemes employ a maximum dictionary size based on the number of available character bits). The easiest way to establish that LZW is not formally equivalent to the assembly index is to look at the scaling of the simplest string case, a repeated single letter (***Figure 3)***. If we use LZW compression, string length, $l$, will grow following:

$$l_{i+1} = l_i + i + 1$$
$$l_1 = 1$$

where $i$ counts the number of compression steps to build up to the next most compressible string of a single letter. This yields a scaling of the longest string length that can be reached by $n$ compression steps (or dictionary construction steps) as

$$l_n = \left(\frac{n(n+1)}{2}\right)$$

Assembly index looks for the shortest path using parts that have already been constructed, and it has previously been shown that this string will grow with assembly steps like[8-10]:

$$l_a = 2^a$$

where we have used the relation of assembly index to string length for single character strings, $a = n - 1$, for $n$ assembly steps. It is clear that the scaling of length versus steps for assembly index and LZW are not equivalent: for $n$ steps in LZW compression, string length scales like $l_n \propto n^2$ (in the limit of large strings) compared to the much faster scaling of $l_a \propto 2^a$ for $a + 1$ steps along an assembly path. Inverting these we find the number of compression steps scales with string length as $n \propto \sqrt{l}$ for LZW schemes, and the number of assembly steps scales like $a \propto log_2 l$. That is, the former is potentially a good compression (as it was designed to be), whereas the latter would be a poor choice if one's goal was a minimum description length (again we note that assembly index was not developed for compression).

While LZW compression is optimal for some contexts it is not optimal in all. For it to be implementable it requires development of a rather sophisticated structure of reading, labeling, and storing data[68]. It is not clear how many complex systems will emergently discover such sophisticated algorithms for data processing. It also depends on how sequences of data are read: data with similar character count, repetition of structure etc., can have very different LZW compressions because these are highly dependent on reading the data from left to right, what characters are at the start of the sequence, and therefore are sensitive to how one chooses to read the data[69].

| | Lempel-Ziv-Welch | | | Assembly | |
|---|---|---|---|---|---|
| Steps | Longest Reachable String | Construction Process | Coding Dictionary | Longest Reachable String | Construction Process |
| 1 | “z” | z | 1, “z” | “z” | z |
| 2 | “zzz” | z+zz | 2, “zz” | “zz” | z+z |
| 3 | “zzzzzz” | z+zz+zzz | 3, “zzz” | “zzzz” | zz+zz |
| 4 | “zzzzzzzzzz” | z+zz+zzz+zzzz | 4, “zzzz” | “zzzzzzzz” | zzzz+zzzz |
| 5 | “zzzzzzzzzzzzzzz” | z+zz+zzz+zzzz+zzzzz | 5, “zzzzz” | “zzzzzzzzzzzzzzzz” | zzzzzzzz+zzzzzzzz |
| | *String Length = 15* | | | *String Length = 16* | |

***Figure 3:*** *The most compressible string of a single letter as represented in $n$ steps of Lempel-Ziv-Welch (LZW) compared with $n$ assembly steps, demonstrating that the scaling of the two algorithms is not equivalent and by extension there can be no formal equivalence between the two.*

Other LZ algorithms such as LZ77 have also been claimed by the same group of authors to be equivalent to, or outperform, assembly index[22–24]. However, these algorithms are always constructed to read data with directionality (e.g. from left to right), with no look-ahead. This feature makes these algorithms efficient – they do not need many passes over the file, or often just one. The Assembly index, on the other hand, look at global optima. It doesn’t just do a look-ahead, it looks over all possible assembly pathways. This is because there is no requirement for speed, but instead to reach an inherent, global, property of the molecule. Sacrificing this feature to make LZ algorithms look more like assembly index calculations therefore removes a central feature of LZ compression schemes, making the direct comparison of these algorithms more of a mathematical exercise in approximating assembly index with a LZ like scheme, then a meaningful statement of exact equivalences between LZ algorithms and assembly index. In the field of computational complexity measures, differences of this degree are sufficient to categorize algorithms as distinct. Indeed, as we will show in the next section computation of assembly index is NP-complete, which puts it in a different computational complexity class than most compression measures, including LZ, meaning it must be formally distinct from them. We note that for molecules, algorithms to compute exact values of assembly index are nontrivial[70], and we do not expect any of the approximate approaches presented in[22] to work in many important cases (this again goes back to the issues of representation of data and labeling, the algorithms the authors implement in[22] will not yield the same results as assembly index in all cases, and would require modification for each new data type for encoding molecular data).

### 3.4 Computation of Assembly Index is NP-Complete

In theoretical computer science, a common way to classify algorithms is by their computational complexity class[27]. If two algorithms are in the same complexity class, they could be equivalent in the difficulty of the computation, although the form of the metrics may not be one to one or indicate the same things about the outputs. However, if two algorithms are found to not be in the same complexity class it is understood they will never be shown to be formally equivalent. We do so by showing that the formal definition of assembly spaces, as originally stated by Marshall et al.[11] and expanded on in the supplement, is NP-complete. It should be noted that thus far in this paper we have focused on strings since they provide simple examples, and while they have not so far been the primary focus of the developers of AT, these are the focus of several works that are critical of the theory. Indeed, AT is more general than the string structures discussed in these criticisms. AT was developed to deal with molecules, and these are much more complicated structures with more complicated attachment rules than string concatenation; thus, we would not expect all properties of string assembly to generalize to the more complex case of molecules. AT, and within it the calculation of the assembly index, covers a wide range of complexity classes. In the proof in the supplement, we give an exact definition of assembly index, and show that a set of string assembly problems are NP-complete. The problem of tiling – whether a shape can be fully covered using a set of tiles – can be included in assembly space and is very similar to assembling a molecule using a set of input molecular structures. Tiling of 2D and 3D shapes in known to be NP-complete[71]. Answering whether a molecule can be assembled out of a set of tiles is simpler than answering what is the minimal number of steps to do so, therefore we know that molecular assembly must be at least NP-complete. Other assembly problems could be simpler. Thus, string assembly is probably NP-hard, whereas the problem of assembling strings of different lengths out of a single letter is equivalent to the shortest addition chain problem, whose exact complexity is unknown, but possibly lower[72].

It is significant to our discussion that assembly index is NP-complete because most compression problems can be solved in polynomial time. Denoted P, this class is defined as the space of all algorithms that can be computed in time bounded by a polynomial function of the size of the input. This makes sense for compression algorithms, because the goal of compression is providing data representations that are easy to compute. Another widely studied class of algorithms live in the class of non-deterministic polynomial time problems, denoted NP, defined as the space of algorithms where a known output has a proof that can be verified in polynomial time by a non-deterministic Turing machine. Non-deterministic Turing machines can be simulated by deterministic Turing machines, but this simulation generally requires exponential time in the worst case. After that, the next class of interest is non-deterministic polynomial time complete problems, denoted NP-complete. NP-complete problems form an equivalence class, defined as the space of algorithms where it is possible to reduce one NP problem to another NP-complete problem in polynomial time. This is relevant because the class of NP-complete problems cannot be mapped to problems in P. Problems in P, however, can be trivially mapped to problems in the NP-complete class, since the problems in P are an easier subset of the NP-complete problems.

The result that computing assembly index for strings is NP-complete has several important consequences. First, NP-complete problems are expected to have an exponential runtime[27] and so we expect algorithms that find the assembly index to have an exponential runtime. Second, proving assembly index is NP-complete means it cannot be strictly equivalent to LZ algorithms, Huffmann coding, or Shannon entropy because these are all in the complexity class P and computing these scales like $O(n)log(n)$, $O(n)$, and $O(n)$ respectively[73,74] as opposed to $exp(n)$. Indeed, any complexity measure that lives in P cannot be strictly equivalent to the assembly index, which is NP-complete. There will be NP-complete problem instances that cannot be mapped to problem instances in P. This result implies our counterexamples are not just specific cases but represent general and fundamental differences between assembly index and these compression algorithms. Comparisons to Huffman coding, Shannon entropy, LZW, and LZ77 can only be approximate comparisons to assembly index and will never represent exact equivalences.

A critical divergence between these compression algorithms and AT lies in interests in optimality. Compression algorithms, while theoretically concerned with finding the shortest possible representation, are more often used in a context where one settles for "good enough" solutions that achieve practical compression ratios quickly. The actual global minimum is not essential for their primary function of efficient data reduction. However, in AT the global minimum (assembly index) is of central interest because it captures minimum causation as an intrinsic physical feature. Note we have not ruled out that computation of the assembly index is equivalent to some NP-complete compression algorithms; for example, optimal, i.e. minimal output length, text-substitution and run-count compression algorithms are NP-complete[74,75]. Shannon entropy is in $O(n)$ where $n$ is the length of the input string, since the algorithm only must iterate over the string once to calculate the frequency of each character. Compression algorithms in the computational complexity class P, such as those reviewed herein, are useful because they are efficient, and in some cases, can be computed with a single pass over the input. Compression algorithms in the NP-complete computation class are intractable for compression and therefore efficient heuristic compression is universally preferred. Assembly index might be considered an optimal compression strategy in this sense since it returns the minimum object reconstruction path, but it is important to recognize the goal of AT is not to provide a formal measure of compression. Finding a formal equivalence with another algorithm that is likewise NP-complete would in no way invalidate the theory but could be useful in providing other ways to compute the assembly index, which might in turn be useful in testing AT's theoretical predictions against experimental data.

Finding that one problem is formally equivalent to another can be useful for calculation, but algorithms are designed to solve specific problem classes and therefore their formulation is usually domain specific. Indeed, all problems in the NP-complete class can be reduced to every other NP-complete problem. *Optimal* Lempel-Ziv is NP-Complete, as is AI and these are therefore reducible to one another. Equivalently, the travelling salesman problem (TSP) is also NP-complete, and therefore the problem of finding the shortest tour of a set of cities separated by roads of various lengths is reducible to optimal Lempel-Ziv and to AI. That is, there is a polynomial time transformation that allows each NP-complete decision problem to simulate the others. However, the equivalence does reduce the usefulness of understanding TSP as movement among cities particularly regarding discussing its correctness. Similarly, the popular Windows video game Minesweeper is NP-complete and therefore there is an efficient mapping from Minesweeper to AI and vice versa, however this is not a particularly helpful reformulation of either Minesweeper or understanding the physics of life to make such a mapping.

Compression algorithms used in practice are heuristics that attempt to approximate optimal compression while being computationally tractable. Similarly, AI requires heuristic approximations for large problems. Heuristics often leverage assumptions about the problem domain to improve their approximations. In the domain of string AI, compression heuristics such as LZ compression may indeed provide useful insights. However, in the case of approximating molecular assembly, relevant to agnostic biosignature detection, the best heuristics will likely leverage domain chemistry knowledge.

## 4 Assembly is a Physical Observable

Many complexity measures can be correlated while not being formally identical. In fact, it should be expected that most computational complexity measures will be correlated with output size, and indeed all measures we have discussed herein, including assembly index, will be (at least weakly) correlated for this reason. In Ozelim et al.[24], the authors claimed correlations between assembly index and other complexity metrics invalidate use of assembly index as a physical observable. However, it is widely recognized that correlation neither implies causation nor equivalence. As a simple illustrative example, consider the two physical laws mathematically represented by the equations $\frac{T^2}{r^3} = \frac{4\pi^2}{Gm}$ and $E = mc^2$. Both are functions of mass with one effective parameter, but each calculates a different physical quantity and in a different context. There is a deep connection between these quantities, but they are built on different effective theories, one for planetary motion and the other for relativistic effects. The two could even correlate perfectly because of their linear dependence on $m$, but this in no way implies the two theories are equivalent or capture the same physics. Likewise, correlation in complexity measures, especially when controlled by overall output size, does not imply equivalence.

### 4.1 Correlated Complexity and the Science of Measurement

In Marshall et al.[11], assembly index was shown to correlate with the number of MS/MS peaks, demonstrating a meaningful measurement of assembly index using mass spectrometry because fragmentation of molecules via mass spectrometry is expected by the theory to probe the structure of the assembly space for molecules via recursive decomposition of molecular fragments. Thus, the measurement using MS/MS maps directly to the ontology of the theory, rendering the correlation physically meaningful. Subsequently, inference of molecular assembly index was shown to lead to convergent results for mass spectrometry, NMR and infrared measurements, again considering molecules as physical objects in AT where these techniques are probing features of the assembly space[12] yielding support for assembly index measurements. This means that the assembly index is invariant with respect to the measurement technique used to read it out and this in turn means that the molecular assembly index is an intrinsic molecular property of a molecule.

The authors in Uthamacumaran et al. [23] and Ozelim et al.[24] show how correlation of different compression measures to assembly index and mass spectrometry data, but do so devoid of ontology and devoid of any grounding of why their reported correlations should be metrologically meaningful. Specifically, they show how the length of string-based representations of molecules (e.g., using SMILES) and compression values for these strings are correlated with the MS/MS peaks in molecular data[24]. They use these kinds of correlations to argue for why assembly index should not be considered as measurable or any better than any other complexity metric that could be used in a similar way[23]. However, their results should be trivially expected based on the scaling cross correlation of complexity measures with output size alone. All complexity measures reviewed herein, including AI, will scale will the size (length) of the output (see e.g., Section 1.3). Thus, if one is not careful to control for string length, comparison across complexity measures will look more correlated with assembly index than these meaningfully are.

This cross correlation is evidenced in[24] where the authors show a plot of correlation between LZW and AI by plotting both scaling with size (Figure 2 in Ozelim et al.[24]). Such correlations are much weaker if one does account for size. As an explicit example, we consider LZW compression and assembly index, comparing the ratio of the two values for strings of fixed length and composition. We considered 10,000 random rearrangements of the string “zbzbczbzbczbzbc" and took the ratio of assembly index to the number of bytes in the compressed string as computed using LZW, see ***Figure 3***. We find that this ratio roughly follows a Gaussian distribution, showing assembly index and LZW these are only weakly correlated with each other when controlling for size. The Pearson correlation between the two is very weak at 0.25. The set of permutations of "*zbzbczbzbczbzbc"* is a small subset of the strings of length 15 constructed from the characters {c,b,z}. We have already shown for the simplest case of strings composed of a single character how LZW scales with length $l$ like $n \propto \sqrt{l}$ while the assembly index scales like $a \propto log_2 l$. This implies that the worst mismatch between LZW and the assembly index will grow at least as fast as $\sqrt{l}/log_2 l$, approximated as $\sqrt{l}$ for large $l$: that is, while both correlate with size, the worst mismatch between the two increases with size. This is also evident in Ozelim et al.[24] where the scatter correlating assembly index and LZW increases with string length.

The cross correlating effect of overall size is expected to be especially prevalent for mean cases (like those in Ozelim et al.[24]), and it is therefore not surprising that any two of the complexity metrics reviewed herein will have mean cases that correlate with size and thus, also with each other. Thus, correlation between any size-correlated complexity metrics with the number of peaks in a mass spectrometer is also not surprising: the number of peaks will itself correlate with molecular size, e.g., molecular weight. Indeed, molecular assembly index is already known to be correlated with molecular weight[76], but there is a large amount of scatter that increases with increasing molecular weight. We should expect that any compression measure will also have mean cases that correlate with MS/MS peaks, but this is not of interest as a physical measurement of complexity. Furthermore, compression algorithms will assign values to molecules that depend on the choice of representation of the data. Theoretical physics proceeds by using specific language in which all statements are intended to carry a physical meaning. In AT, this is done by selecting a scheme for system representation and specifying a set of operations that correspond to physically realizable transformations. For molecular AT, this choice is well-motivated by knowledge of covalent chemistry[77]. For new domains of application more work is to be done to ground equally well-motivated choices for construction of the assembly space in empirical measurement[76]. Computational complexity methods face serious ambiguities as tools for physics because they suffer from an arbitrary choice of universal computer, whereas AT resolves this ambiguity through metrology.

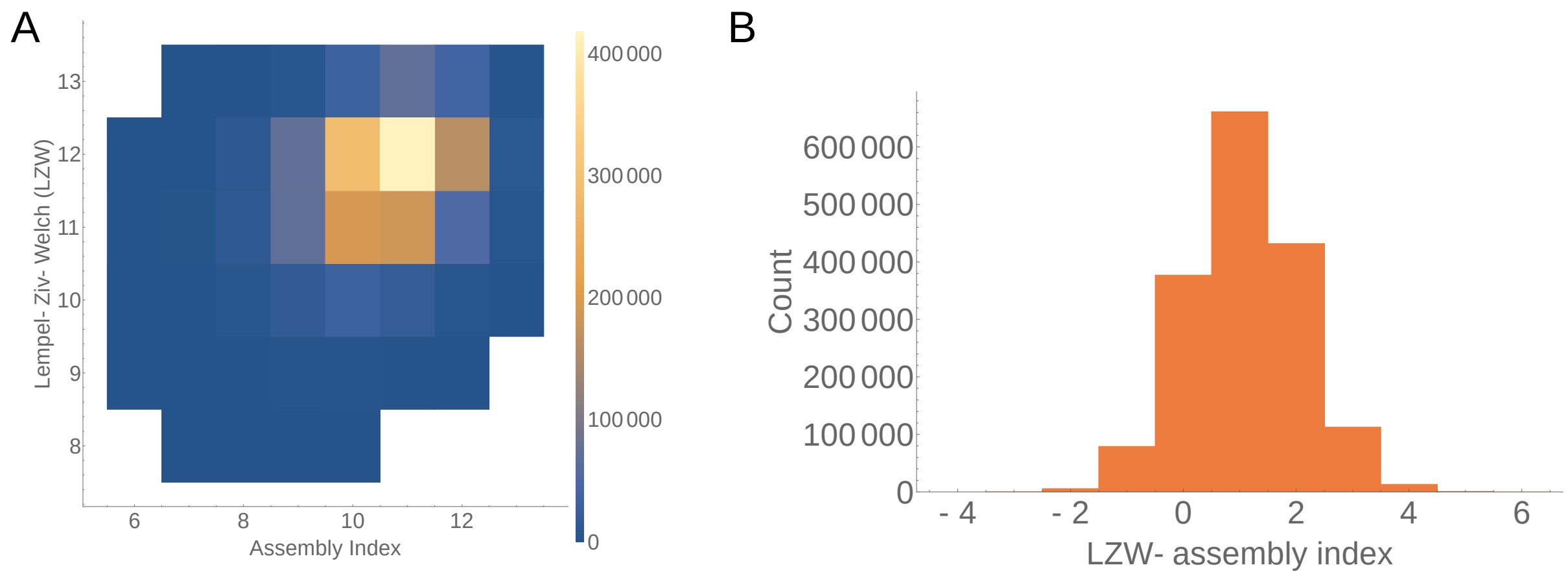

***Figure 3:*** *Comparison of the assembly index, a, and the Lempel-Ziv-Welch compression for every permuation of "zbzbczbzbczbzbc". A) The density map of a compared with LZW where the two have a weak Pearson correlation of 0.25. B) The histogram of the difference between a and the length of LZW (given as LZW – assembly index) showing an approximate Gaussian distribution. These results confirm that when controlling for size, assembly index and LZW compression are weakly correlated.*

The measures the authors of refs 22-24, introduce all suffer from ambiguity in the choice of data representation, analogous to the ambiguity in choosing a universal computer for $K$ (see Section 1.1). They will lack an absolute ordering for complexity that would accurately delineate the boundary between non-life and life, obfuscating the distinction that AT formalizes. Indeed, different data representations of the same molecule, e.g., in the form of SMILES[78], or INCHI[79], graph-based representations, or even mass spectrometry data files should be expected to yield *quantitatively* different results for the computed compressibility of the molecular data. Thus, a general correlative trend will always be established, but the values assigned to specific molecules will depend on how one chooses to represent the data (e.g., this characterizes what is observed Figure 7 in ref 24). Using these correlations would provide a very poor basis for measurement because it should be expected that the same molecule will fall in different locations on such correlation plots depending on choices made representing the molecular data. That is, we should expect correlation to always be established, but the structure underlying where specific molecules lie to be representation dependent. This is untenable for life detection, because as we have already pointed out, one could choose any representation of the data and its compression that would represent any claim one might wish to make about the relative complexity of the molecules (Section 1.1), thus rendering the correlations useless without an underlying ontology forcing a choice of representation (what physics does in selecting meaningful languages of description).

Confusing the cross correlation of mean values of data compression and output size[24], with an empirically grounded metric represents a fundamental misunderstanding of the science of measurement and its historical development[39]. Measurements are constructed by careful mapping between the phenomenology of a theory and what a measuring device is probing. The formalization of assembly index was inspired by the physical measurement of recursive molecular structure first interrogated via mass spectrometry[11], and later confirmed by NMR and infrared techniques[12]. The assembly index is intended to be independent of the descriptive labels a human might want to put on chemical data, because the measure is embedded in a standardized measurement scheme built not solely on correlation but an understanding of how the structure of the assembly index manifests in different modalities of interrogating covalently bonded molecules[76]. The assembly index is therefore metrologically relevant, while correlation with compression metrics has not been shown to be.

### 4.2 Physical Complexity is Measurable in Assembly Theory

Many considerations about the physical nature of the process of computation, as well as the physical instantiation of outputs of algorithmic procedures, are not accounted for by the theory of computation[27]. This has motivated several researchers to develop related complexity measures that might capture more physically relevant details. We now discuss these measures, and why AT represents a significant advance in the development of a physics of complexity, measurable in the laboratory. Originally developed by Bennett[80], logical depth was inspired by universal computation but with a goal to quantify achieved complexity rather than potential complexity. Logical depth is distinct from algorithmic complexity because it quantifies complexity based on the work a computer must do to produce a given output, rather than an abstracted notion of program size. Formally, the logical depth of a string is defined as the execution time, or number of machine cycles, it takes a computer to calculate the output from the shortest possible program. In general, these yield different results: random objects have little depth despite having potentially long descriptions, while structured objects like repeating strings are expected to have a depth more coincident with the difficulty of the process that generated them.

As an explicit example, if we compare **Program 1** and **Program 2** from Section 1.1, the less complex string $x_2$ would have a greater logical depth because the program must execute 72 machine cycles ($x \rightarrow x + 1, printxmod36$) whereas the random string $x_1$ only requires one cycle to print. Here we have an explicit example where a difference in the ontology of complexity leads to quantitative differences. For algorithmic complexity "complex" is determined by the size of the minimum description, and for logical depth "complex" is determined by the run-time of the minimum description. Logical depth has an advantage in applications where it is necessary to distinguish complexity due to randomness as distinct from meaningful computational work. However, like algorithmic complexity it is dependent on the choice of language to describe the output and in general the shortest program is not computable.

Thermodynamic depth was introduced by Lloyd and Pagels as a measure of macroscopic complexity[81] and is closely related to, but differs from, logical depth. Thermodynamic depth captures the number of steps a system passes through to reach the current state. The goal of thermodynamic depth is to quantify complexity as a physical property of a state, noting that complexity should be a function of the process, or assembly routine, that brought the state into existence (and not just a description of the state)[38]. We have now arrived at the first comparison complexity measure reviewed herein which deals explicitly with physical states rather than computational programs and their outputs. Thus, this is the first candidate, besides assembly index that is explicitly about physics and what can be experimentally determined. Lloyd and Pagels tried to empirically ground their measure of physical complexity by defining it as a continuous function of the probabilities of the experimentally defined trajectories that can result in a macrostate, $d$, which might be observed in a lab. The thermodynamic depth is thus defined as:

Eq 4.
$$D(d) = -kln(p_i)$$

where $p_i$ is the probability, the system arrived at state $d$ by the $i$th observed trajectory and $k$ is an arbitrary constant. Thermodynamic depth will depend on what experiment is done, e.g. via empirically observed frequencies, $p_i$, of the observed paths. Its value can therefore differ for the same macrostate simply because the experiment was prepared differently (or more experiments were done). This introduces subjectivity in its definition[82].

In a chemical system the thermodynamic depth would depend on the probability of the reaction pathways that produced it[81]. This is untenable as a complexity metric for life detection because the same molecules will have a different thermodynamic depth in different contexts (where the reactions to generate the molecule would differ). This measure is therefore not standardizable across different planetary environments or different experiments. Also, one cannot infer thermodynamic depth from an observation of a complex chemical system at just one point in time but must instead know the full history of its formation: this information is almost never available for many scientific questions of interest. These challenges are exacerbated by the astronomical size of chemical space, where counting all possible reactions, pathways and molecules is not computationally possible[83–85]. The assembly index, by contrast, is independent of the path that produced the object(s), rather it depends only on the minimum path that could have produced the object, and therefore can be measured without knowing formation history. It will yield the same value for the same physical object(s) anywhere in the universe we might observe them. The length of the minimum recursively constructed path to make an object is a unique value for every object (even in cases where there may be degeneracy in minimum paths for the object's construction, the minimum path size is still a unique value and measurable).

**5. Copy Number**

Falsification of any physical or chemical theory should come from experimental and observational tests of the explanations it provides, and the predictions it makes. AT was developed by critically addressing the question of how complex a detectable molecule must be before we know it was not formed by a random or undirected process. Exploring this idea allowed the concept of causal or contingent control, at the molecular level, to be formalized as the explanation of how highly complex objects are produced in abundance. AT as a theory is still in its infancy, with many open questions remaining about the mechanism for selection, and the transition in Assembly from the abiotic world to the biotic world. What is clear is that AT is poised to help uncover how selection, as a driving force, allowed the evolution of evolution in a manner that is testable with standardized measurements. As the theory develops it will be important to explore its foundations. In this regard it is worth noting that any good scientific theory should be "hard to vary"[86], especially as concrete predictions are made that can be experimentally or conceptually tested and built on. This is one reason the assembly index having an exact value, not dependent on data representation, is critical in its role as a foundation of the theoretical framework of AT.

Critical to the conceptual shift presented by AT is the central role of copy number, which so far, we have not addressed here. Criticisms of AT as redundant or not new, have ignored the bulk of the theory, including the structure of the assembly equation and its role in formalizing selection and the observed assembly space as a necessarily contingent structure[22–25,87]. In AT, objects must be found in multiple copies. This means in applications of the theory it is not meaningful to talk about the assembly index of objects independent of any considerations of how to measure and count identical copies of them (where "identical" is defined within measurement uncertainty). Whilst this may seem an *ad hoc* requirement, it is a critical feature of AT's ability to uniquely identify objects as products of selection and evolution: the conjecture of the theory is high Assembly configurations of matter (with abundant, large assembly index objects) only occur via evolution and selection. Note in Eq. 1 if $n_i = 1$, then the contribution of object $i$ to the total value of $A$ is 0. A recent hypothesized falsification of AT did not address the conceptual foundations of the theory accurately[87], because it applied the theory to idealized mathematical representations of unit cells, where in real materials these are not possible to be observed in identical copies due to defects[76].

In a recent discussion of AT, it was stated that AT cannot account for unanticipated biases, e.g., that AT cannot distinguish whether the bias is natural selection or something else that might favor production of an object given the "rule-based world" that AT is modeling[25]. However, this criticism is made without looking at the empirical data and seemingly without understanding the ontology of the assembly index (in this same work the author claims assembly index is a "special case" of algorithmic complexity). AT is a link between the inorganic world and the biological world and shows that a much more generalized mechanism of selection must have acted before genetic selection appeared. That is, AT makes explicit that selection is a more general phenomenon than that found just in biology. Our central claim is that the combinatorial space is so large, that some configurations simply cannot exist without "systematic bias" stored in a memory, what we call "selection" which will include modern genetic evolution, but also many other more general modes of selection. So far, the central claim holds up to empirical tests, as presented previously[11] and those are not explored in this criticism. Also, the author does not provide any empirical evidence to support the argument that the mundane deviations from expected outcomes they cite would falsify the central hypotheses of AT by producing arbitrarily high Assembly configurations of matter in abiotic settings.

Another interesting critique presented by the same author is related to how AT does not explicitly address function[25]. However, a goal of AT is to measure evolved complexity *without* referencing function. This is explicit, as it avoids the issue of predefining all possible functions of a given object where there will always be some unanticipated context in which the object might acquire new function[88]. Indeed, it is unclear if the possible uses of a complex object are finite or (at least asymptotically) infinite[89]. If one has N objects there are potentially $\sim N^N$ possible combinatorial interactions meaning the number of possible interactions (functions) trends super-exponentially faster toward infinity than the number of objects does as the system size grows. However, while the number of functions an object can perform is not pre-statable [90], the number of functions that could have constructed any observed object is always **finite** and **well-defined**. This is exactly what underlies the definition of the assembly index and the assembly space: AT explains how an object found in high copy number is constructed by functions, rather than what function it has, specifically because this is finite and measurable, as opposed to undefined. As an explicit example of the utility of this approach one can consider the long-standing paradoxical question of how to tell if a given sequence of DNA is "functional" or not, without referring to the context of the cell[91]. AT would instead ask whether the abundant DNA found in the cell was evidence of selection or not.

Well-developed theories often lead to counterintuitive reformulations of our understanding of phenomenon that depart from our colloquial understanding of them, reframing the question. It may be the case that a formal theory of life will require abandoning our biases about the role and nature of function, to learn something deeper by building abstractions that tie more closely to what empirical data we can objectively measure. Whether AT will continue to pass tests of its empirical validation or be falsified will be determined by future work of the scientific community. But one should not claim a theory invalid because it does not conform to prior expectations of what explanations we should find. Instead, we must always be open as scientists to theories we can rigorously test, and the explanations they provide.

## 6. Conclusions

We have shown that the assembly index is not identical to measures of computational complexity and compression, and we have discussed how the phenomenology of assembly theory as a theory of physics rooted in measurement distinguishes it from computational complexity theory. Assembly theory was designed to explore how the evolution of evolution was possible before formalized genetics. That is, AT aims to elucidate how selection produces the machinery of evolution. One of the challenges with traditional computational complexity theory as an approach to physics and biology is that it has been applied to quantify the amount of complexity found in the natural world using an ontology that may not be meaningful in this context, using metrics that cannot be meaningful mapped directly to physical measurement. Whereas the other measures described herein are in some cases undecidable, and in all cases dependent on choices of data representation, AT provides a formalization dependent on the physics that generated the object with a grounding in metrology. Furthermore, AT makes predictions falsifiable through empirical observations. One prediction is that objects experimentally inferred to have high assembly index and copy number are uniquely associated with life. Falsifiability is not unique to AT as a scientific theory. What is unique is the ability to use readily available biochemistry lab equipment to obtain fragmentation patterns and copy-number data that map to the AT framework, making it the first theory tailored to life detection that is amenable to experimental validation.

## Acknowledgements

We thank Abhishek Sharma, Start Marshall, Keith Patarroyo, Gage Seibert and Cole Mathis for ideas and comments, Ian Seet for efficient algorithms to compute string assembly, and Louie Slocombe for help making Figure 1. We thank the Santa Fe Institute and the NSF ‘RCN for Exploration of Life’s Origins’ (NSF grant no. 1745355) for supporting an ongoing set of working groups that have facilitated this work. We also thank the John Templeton Foundation (grant nos. 61184 and 62231), the Engineering and Physical Sciences Research Council (EPSRC) (grant nos. EP/L023652/1, EP/R01308X/1, EP/S019472/1 and EP/P00153X/1), the Breakthrough Prize

Foundation and NASA (Agnostic Biosignatures award no. 80NSSC18K1140), MINECO (project CTQ2017-87392-P) and the European Research Council (ERC) (project 670467 SMART-POM) for financial support.

# Assembly Theory and its Relationship with Computational Complexity
# (Supplementary Material)

Christopher P. Kempes, Michael Lachmann, Andrew Iannaccone, G. Matthew Fricke, M. Redwan Chowdhury, Sara I. Walker, Leroy Cronin

## 1 Proof That the Assembly Index Problem Is NP-Complete

We prove that the assembly index problem (AssemblyIndex) is **NP-Complete**[1] and is therefore computable but intractable. Large instances of problems in this class require heuristic approaches that provide approximate solutions.

Computer scientists arrange problems into classes defined by the computational resources needed to solve those problems (see Arora and Barak [1] for a thorough discussion of this area of computer science). Some important classes are **P**, **NP**, **NP-Complete**, and **NP-Hard** which describe the amount of *time* required to solve the problems they contain. Problems in **P** can be solved by a deterministic Turing Machine [10] (DTM)[2] in time bounded by a polynomial function of the size of the input. Problems in **P** are notionally efficient or tractable. Solutions to problems in **NP** can be found in exponential time by a DTM,[3] **NP-Hard** problems take at least as much time to solve as any problem in **NP**, and are considered intractable.[4] **NP-Complete**, the class to which we demonstrate AssemblyIndex belongs, contains those problems which take no more than exponential time but strictly more than polynomial time.

Some polynomial time algorithms can grow quickly (e.g. the Picture Hanging Problem is in **P** with $O(n^{43737})$ [3] and some superpolynimial problems can grow slowly, e.g., Planar Hamiltonian Circuit is **NP-Hard** but with time complexity $O\left(C^{\sqrt{n}}\right)$ [4]. However, for sufficiently large problem sizes **NP-Hard** problems will always take longer to solve than those in **P**.

This has practical implications for calculating AssemblyIndex. **NP-Complete** problems require more than polynomial time in the size of the problem instance. As problem sizes become sufficiently large the resources required to solve them become unmanageable and require heuristic approaches that may be faster but cannot guarantee exact solutions. Membership in these classes help to differentiate algorithmic classes from one another. For example, a problem known to be in **P** cannot be equivalent to another algorithm that is **NP-Complete**, unless the computational classes were to collapse.[5]

We prove that the AssemblyIndex problem is a member of the classes **NP** and **NP-Hard**, and is therefore **NP-Complete**. The Vertex Cover problem (VertexCover) is known to be **NP-Hard** [5]. By performing a Karp reduction of VertexCover to the calculation of AssemblyIndex, we show that AssemblyIndex is also **NP-Hard**. In other words, it is demonstrated that any instance of VertexCover can be solved by reformulating it as an AssemblyIndex problem and that the reformation can be performed in polynomial time. Calculating the AssemblyIndex is therefore at least as hard as solving VertexCover, since Vertex Cover problems are a subset of AssemblyIndex problems. To demonstrate that AssemblyIndex is not only **NP-Hard** but also **NP-Complete** we prove that AssemblyIndex is a member of **NP**.

The existence of an efficient algorithm for solving AssemblyIndex requires that **P**=**NP**. On the other hand, AssemblyIndex can be solved in no more than exponential time, and is therefore distinct from harder problems such as the Halting Problem and Kolmogorov Complexity which have no solution at all for certain inputs. AssemblyIndex lies in the class of solvable problems that require heuristic solutions for all but the most trivial instances.

[1] A problem is a member of a complexity class, e.g. AssemblyIndex $\in$ **NP-Hard** but following convention we will often write AssemblyIndex *is* **NP-Hard**

[2] Deterministic Turing Machines are used to reason about physical computers.

[3] That is, in non-deterministic polynomial (NP) time by a non-deterministic Turing Machine. Non-deterministic Turing Machines are non-physical models of computation.

[4] **NP-Hard** problems require superpolynomial time to solve.

[5] Strictly speaking it has not been proven that there are no polynomial solutions to **NP** problems. This is the famous **P** $\neq$ **NP** conjecture.

**Definition 1** (Karp Reduction)**.** Given decision problems $A$ and $B$, $A$ is Karp reducible to $B$ if there a function $f$ such that 1) for every instance $x$ of problem $A$, $x$ is a *yes* instance of $A$ if and only if $f(x)$ is a *yes* instance of $B$ and 2) $f$ can be computed in polynomial time [5]. We denote the Karp reduction of problem A to problem B with $A \leq_m^p B$.

**Definition 2** (NP-Complete)**.** For problem $p$, $p \in$ NP-Complete if $p \in \text{NP} \wedge \forall x \in \text{NP} x \leq_m^p p$. That is, for a problem to be in the class of **NP-Complete** problems it must be in the class of **NP** problems but also as hard as every other **NP-Complete** problem [2]. Equivalently NP-complete = NP $\cap$ NP-Hard.

**Definition 3** (Vertex Cover)**.** Given a graph $G = (V, E)$, the vertex cover decision problem asks if whether there is a subset $C \subset V$ with $|C| \leq k$ such that every edge in $E$ includes at least one vertex in $C$ [5].

The proof is divided into four sections: In Section 2 we define the terms used throughout the proof, in Section 3 we prove that AssemblyIndex is in **NP**, and that it is **NP-Hard** in Section 4. Finally we bring everything together in Section 5 to prove that AssemblyIndex is **NP-Complete**.

## 2 Formalising Assembly Spaces

We use the AssemblyIndex definitions from Marshall et al. [6] where the problem was originally introduced, but reformulate them for for clarity in our context. For example, we use graph theory terms rather those of path algebra.[6] We also introduce some novel related definitions and notation, and we also restate some definitions from graph theory for the convenience of the reader.

To mathematically specify a set of assembly objects and the ways in which they can be assembled, we use a directed graph with labeled edges. The vertices of the graph represent the assembly objects, and the directed edges represent allowed assembly operations.

**Definition 4** (Edge-labeling Map)**.** Given a directed graph $\Gamma$ with vertices $V(\Gamma)$ and edges $E(\Gamma)$, an *edge-labeling map* is a function $\phi : E(\Gamma) \to V(\Gamma)$ that assigns a vertex to every edge of the graph.

[6] For example, we refer to directed graphs rather than quivers and source vertices rather than the minimal set.

**Definition 5** (Edge Bracket Notation)**.** In the context of a directed graph with vertices $x$ and $z$, we use $e = [x, z]$ to mean that $e$ is a directed edge from vertex $x$ to vertex $z$.

**Definition 6** (Reachable)**.** A vertex $z$ in a directed graph $\Gamma$ is **reachable** from vertex $x$ if, and only if, there is a path along directed edges in $\Gamma$ from $x$ to $z$. Every vertex can also reach itself along a zero length edge, i.e. the relation is reflexive.

**Definition 7** (Source Vertex Set: $\text{Src}(\Gamma)$)**.** A vertex $x$ in a directed graph $\Gamma$ is a **source vertex** in $\Gamma$ if it is reachable only from itself, i.e. it has indegree 0. We use $\text{Src}(\Gamma)$ to denote the set of source vertices in $\Gamma$.[7]

**Definition 8** (Assembly Space)**.** An **assembly space** $\Omega = (\Gamma, \phi)$ is an acyclic directed graph[8] $\Gamma$ together with an edge-labeling map $\phi$ for which:

1. $\text{Src}(\Gamma)$ is finite and non-empty,
2. Every vertex in $\Gamma$ is reachable from a vertex in $\text{Src}(\Gamma)$, and
3. if an edge $e = [x, z] \in E(\Gamma)$ and $\phi(e) = y$, then there exists an edge $e' = [y, z] \in E(\Gamma)$ for which $\phi(e') = x$

This definition is much complex than it appears at first glance. The first two criteria are trivial for any graph $\Gamma$ that is finite and non-empty, and the third criterion can be roughly understood as "the order in which a pair of objects is combined does not matter." We have used this definition to remain consistent with existing literature, but as we will not be using any empty or infinite assembly spaces in this appendix, we recommend the reader internalize the simpler Definition 11 of a ***finite* assembly space** for our purposes.

**Definition 9** (Assembly Object)**.** If $\Omega = (\Gamma, \phi)$ is an assembly space, we refer to the vertices of the directed graph $\Gamma$ as its *assembly objects*, or just its *objects*. That is, we define

$$V(\Omega) \triangleq V(\Gamma) \tag{1}$$

The assembly objects of the assembly space $\Omega$ *are* the vertices of its directed graph $\Gamma$.

**Definition 10** (Assembly Operations)**.** If $\Omega = (\Gamma, \phi)$ is an assembly space, and there is an edge $e = [x, z] \in E(\Gamma)$ with $\phi(e) = y$, we call this labeled edge an *assembly operation* (or simply *operation*) and denote it by

$$[x + y \to z] \tag{2}$$

We also say that this assembly operation in *allowed* in $\Omega$, and that this operation *assembles $z$ from $x$ and*

[7] Marshall et al. 2019 [6] instead use $\min(\Gamma)$ for the set $\text{Src}(\Gamma)$ and say the vertices in this set are *minimal* in $\Gamma$.

[8] Marshall et al. use the equivalent term "quiver".

$y$. We define $E(\Omega)$ as the set of all allowed assembly operations in $\Omega$, so we will also write

$$[x + y \to z] \in E(\Omega) \tag{3}$$

Alternatively, if there is no such edge $e = [x, z] \in E(\Gamma)$ with $\phi(e) = y$, we say that $[x + y \to z]$ is *not allowed* in $\Omega$, and we write

$$[x + y \to z] \notin E(\Omega) \tag{4}$$

This allows us to rewrite third criterion of Definition 8 as:

$$[x + y \to z] \in E(\Omega) \Longleftrightarrow [y + x \to z] \in E(\Omega) \tag{5}$$

It is possible to further simplify Definition 8 if we are only interested in assembly spaces that have a finite, non-zero number of assembly objects.

**Definition 11** (Finite Assembly Space)**.** A *finite assembly space* $\Omega = (\Gamma, \phi)$ is a finite, non-empty, acyclic directed graph $\Gamma$ together with an edge-labeling map $\phi$ for which:

$$[x + y \to z] \in E(\Omega) \iff [y + x \to z] \in E(\Omega) \tag{6}$$

**Lemma 1.** *Every finite assembly space is an assembly space.*

*Proof.* Let a finite assembly space $\Omega = (\Gamma, \phi)$ be given. The Definition 8.3 criterion is equivalent to Equation (6), so we only need to verify the criteria of Definition 8.1 and Definition 8.2:

1. $\mathrm{Src}(\Gamma) \neq \emptyset$ and is finite:
   (a) $\mathrm{Src}(\Gamma)$ is finite because $\mathrm{Src}(\Gamma) \subset V(\Gamma)$, and $V(\Gamma)$ is finite.
   (b) To show that $\mathrm{Src}(\Gamma) \neq \emptyset$: choose any vertex $v \in \Gamma$ and apply Algorithm 1, which returns a path from some source vertex $u$ to $v$ (with the possibility that $u = v$). Note that Algorithm 1 cannot loop indefinitely because the graph is finite and acyclic. Because $u \in \mathrm{Src}(\Gamma)$, $\mathrm{Src}(\Gamma) \neq \emptyset$.
2. For any arbitrary $v \in V(\Gamma)$, Algorithm 1 returns the exact path from some vertex $u \in \mathrm{Src}(\Gamma)$ to $v$, so $v$ is reachable from $u$.

□

**Definition 12** (Assembly Subspace)**.** Given two assembly spaces $\Omega = (\Gamma, \phi)$ and $\Omega' = (\Gamma', \phi')$, we say that $\Omega'$ is a *subspace* of $\Omega$ if $\Gamma'$ is a subgraph of $\Gamma$ and the edge-labeling functions $\phi$ and $\phi'$ are consistent. That is, if

**Algorithm 1** Find a path from some source vertex to given vertex v

**Require:** Directed Acyclic Graph $G = (V, E)$, target vertex $v$
**Ensure:** Returns a path from any source vertex $u$ to $v$, or **nil** if no such path exists
**function** PathToSourceVertex(G, v)
  **Input:** Directed Acyclic Graph $G = (V, E)$, target vertex $v$
  **Output:** A path from any vertex $u$ with only outgoing edges to $v$, or **nil**
  **visited** ← **False** for all $x \in V$
  **function** DFS(x)
    **visited**[x] ← **True**
    **for each** $y$ such that $[y, x] \in E$ **do**
      **if not visited**[y] **then**
        **path** ← DFS(y)
        **if path** ≠ **nil then**
          **return path** + $[y]$
    **if** $\forall w \in V,\ (w, x) \notin E$ **then** ▷ Check if $x$ has only outgoing edges
      **return** $[x]$
    **return nil**
  **return** DFS(v)

1. $V(\Gamma') \subset V(\Gamma)$
2. $E(\Gamma') \subset E(\Gamma)$
3. $\phi'(e) = \phi(e) \forall e \in E(\Gamma')$.

We write this assembly subspace relationship as $\Omega' \subset \Omega$.

**Definition 13** (Basis)**.** We say an assembly object $x$ in an assembly space $\Omega = (\Gamma, \phi)$ is a **basic object** in $\Omega$ if $x$ is a source vertex in $\Gamma$. We call the set all basic objects the **basis** of the assembly space, and we denote it $B_\Omega$. That is,

$$B_\Omega = \mathrm{Src}(\Gamma) \tag{7}$$

**Definition 14** (Composite Object)**.** We say $x$ in is a *composite object* in an assembly space $\Omega$ if $x \in V(\Omega)$ and $x \notin B_\Omega$.

Note that an object $x$ in an assembly space $\Omega$ is a composite object if and only if there is an allowed assembly operation in $E(\Omega)$ that assembles it.

**Definition 15** (Rooted Subspace)**.** We say that a subspace $\Omega'$ of an assembly space $\Omega$ is **rooted in** $\Omega$ if $B_{\Omega'} \subset B_\Omega$. We also say that $\Omega'$ is a **rooted subspace** of $\Omega$.

Conceptually, a rooted subspace is one in which every object can be assembled from the basic objects of the original space.

## 3 The Assembly Index Decision Problem

We follow the formal definition established in Marshall et al. 2019 [6].[9]

**Definition 16** (AssemblyIndex)**.** For an assembly object $x$ in an assembly space $\Omega$, we say the **ASSEMBLY INDEX of x in** $\Omega$ is the minimum number of composite objects in any rooted assembly subspace of $\Omega$ that contains $x$. We denote this number $a_x$.

Complexity classes are defined in terms of decision problems rather than optimisation problems. For the remainder of this proof ASSEMBLYINDEX refers to the decision problem unless otherwise specified.

**Definition 17** (THE ASSEMBLYINDEX Problem)**.** The **ASSEMBLYINDEX *decision* problem** is: Given an assembly space $\Omega = (\Gamma, \phi)$ containing a object $x$, is there a rooted subspace of $\Omega$ that contains $x$ and has no more than $k$ composite objects? The input size of the problem is $N = |\Gamma|$.

The *optimisation* version of the problem asks instead for the minimum number of composite objects in any rooted subspace of $\Omega$ that contains $x$. This optimisation can be solved by calling the decision version in Definition 17 as a subroutine a polynomial number of times. For example, by using an $O(n)$ linear search for a $k$ where the answer is *yes*. Therefore, showing that the decision problem is **NP** and **NP-Hard** classifies the optimisation problem as well.

Definition 17 is the general version of the ASSEMBLYINDEX problem– one that is relevant to any type of assembly object: molecules, 2D pixel art, strings, etc. We can also define more limited versions of the problem that are restricted to these specific domains.

Because there has been particular interest in Assembly Theory applied specifically to string objects, we will define the STRINGASSEMBLYINDEX problem in Section 4.4.1, and we will show that both the general problem and the restricted string version are **NP-Complete**.

**Lemma 2** (ASSEMBLYINDEX is in **NP**)**.** *A proposed solution to an* ASSEMBLYINDEX *decision problem (certificate) can be checked in polynomial time.*

*Proof.* Let an assembly space $\Omega = (\Gamma, \phi)$ be given, along with an assembly object $x$ and an integer $k$. Also let a certificate $\Omega' = (\Gamma', \phi')$ be given. We will show that this candidate solution can be checked in $O(N^2)$ time, where $N = |\Gamma|$ is the input size of the ASSEMBLYINDEX problem.

First, we must check whether $\Omega'$ is an assembly space at all. We will presume that the directed graph $\Gamma'$ is indeed finite, and we can check that it is non-empty in $O(1)$ time. To check whether $\Gamma'$ is acyclic, we perform a depth-first search of $\Gamma'$ starting from each of its vertices in turn. If such a search ever reaches a vertex that is already on the recursion stack (a vertex already visited along the current DFS route), then we have found a cycle in $\Gamma'$, and the certificate fails. If this never happens, then our $|V(\Gamma)|$ depth-first searches will complete in $O(|\Gamma|^2)$=$O(N^2)$ time, since a single DFS on an acyclic graph $\Gamma$ can be performed in $O(|\Gamma|)$ time. For $\Omega'$ to satisfy the definition of a finite assembly space (Definition 11), it remains only to verify that

$$[x + y \to z] \in E(\Omega) \iff [y + x \to z] \in E(\Omega). \quad (8)$$

To do so, we can perform a nested iteration over the edges of $\Gamma'$. For each edge $e = [x, z] \in E(\Gamma')$ with $\phi'(e) = y$, we search for an edge $e' = [y, z] \in E(\Gamma')$ with $\phi'(e') = x$. If we fail to find any such edge $e'$, then the certificate fails. If not, then this nested iteration will require $O(N^2)$ time, and we will know that $\Omega'$ is an assembly space.

Next, we will check that $V(\Gamma') \subset V(\Gamma)$, that $E(\Gamma') \subset E(\Gamma)$, and that $\phi$ and $\phi'$ label the shared edges consistently. Each check can be done by nested iteration in $O(N^2)$ time. If all three checks pass, we will know that $\Omega' \subset \Omega$.

Next, we will determine $\mathrm{Src}(\Gamma)$. For each vertex $v \in V(\Gamma)$, we search for an edge in $E(\Gamma)$ whose terminal vertex is $v$. If no such edge is found, then $v$ is reachable only from itself, so $v \in \mathrm{Src}(\Gamma)$. Repeating this for each vertex will fully determine $\mathrm{Src}(\Gamma) = B_\Omega$ in $O(N^2)$ time. We can then do the same for $\mathrm{Src}(\Gamma') = B_{\Omega'}$. We then compare the two sets to determine whether $B_{\Omega'} \subset B_\Omega$, which can be done by nested iteration in $O(N^2)$ time. At this point, we will know whether $\Gamma'$ is rooted in $\Gamma$.

All that remains is to check whether $x \in V(\Omega')$ and to calculate $|V(\Omega')| - |B_{\Omega'}|$, the number of composite objects in $\Omega'$, and compare it to $k$. At this point, we will have fully checked the certificate in $O(N^2)$ time. □

We now know that the ASSEMBLYINDEX problem is in NP. We will prove that it is also **NP-Hard** indirectly, by showing there is a subclass of ASSEMBLYINDEX problems that is **NP-Hard**.

[9] We have introduced the concept of a "composite" object and omitted the Marshall et al. definition of "augmented cardinality," but the definitions are mathematically equivalent.

## 4 AssemblyIndex is NP-Hard

We define StringAssemblyIndex to be a subclass of AssemblyIndex problems in which all of the assembly objects are strings, and all of the assembly operations are string concatenation. To obey the definition of assembly space, we use a modified definition of concatenation in which assembly operations are commutative. By proving that this subclass is **NP-Hard** we also prove that the superclass is **NP-Hard**.

### 4.1 The String Assembly Index Decision Problem

**Definition 18** (String Product Notation)**.** When $x$ and $y$ are strings, we use $xy$ to denote their concatenation. We use power and product notation similarly, e.g.

$$\prod_{i=1}^{2} x^i y = (xy)(xxy) = xyxxy \tag{9}$$

**Definition 19** (String Assembly Space)**.** We say an assembly space $\Omega$ is a **string assembly space** if all of assembly objects in $V(\Omega)$ are strings, and if

$$[x + y \to z] \in E(\Omega) \implies z = xy \vee z = yx. \tag{10}$$

That is, if any two objects $x, y \in V(\Omega)$ can be assembled, then the object they assemble must be one of their concatenations, either $xy$ or $yx$.

It is it important to note, however, that since this concatenation operation is commutative it is not what we may normally think of as concatenation where $[x + y \to xy]$ and $[y + x \to yx]$. In particular, $[x + y \to xy] \in E(\Omega)$ does not imply $[y + x \to yx] \in E(\Omega)$, as might be expected for string concatenation. In fact, it actually implies $[y + x \to xy] \in E(\Omega)$, because Definition 11 requires every assembly operation to be commutative. Whether the non-commutative concatenation version of string "assembly" is NP-complete is an open question.[10]

Also note that Definition 19 does not imply that all possible string concatenations are allowed assembly operations in $\Omega$. It isn't even required for the concatenation of any two strings to be an assembly object in $\Omega$. Any given concatenation might be an allowed assembly operation or might not be, depending on whether such an edge exists in the assembly space's directed graph. In particular:

$$x, y \in V(\Omega) \not\Rightarrow xy \in V(\Omega) \tag{11}$$

$$x, y, xy \in V(\Omega) \not\Rightarrow [x + y \to xy] \in E(\Omega) \tag{12}$$

$$[x + y \to xy] \in E(\Omega) \tag{13}$$

$$\not\Rightarrow [y + x \to yx] \in E(\Omega) \tag{14}$$

**Definition 20** (StringAssemblyIndex Problem)**.** The **StringAssemblyIndex decision problem** is: Given a string assembly space $\Omega$ containing a string $x$, is there a rooted subspace of $\Omega$ that contains $x$ and has no more than $k$ composite objects?

The StringAssemblyIndex problem is a specific case of the AssemblyIndex problem, so if we can show that it is **NP-Hard**, it will immediately follow that the general problem is also **NP-Hard**.

### 4.2 StringAssemblyIndex is NP-Hard

**Lemma 3** (StringAssemblyIndex is **NP-Hard**)**.** *Any decision problem in* **NP** *can be reduced to a* StringAssemblyIndex *decision problem in polynomial time.*

*Proof.* We show that VertexCover $\leq_m^p$ StringAssemblyIndex (Definition 1). That is, given any VertexCover problem (Definition 3), i.e. any finite undirected graph $G = (V, E)$ and any positive integer $k$, we can solve the problem by reducing it to an instance of StringAssemblyIndex in a polynomial number of steps. This is accomplished by defining a string assembly space $\Omega = (\Gamma, \phi)$, a string $x$, and an integer $k'$ so that the resulting StringAssemblyIndex decision problem will always have the same answer as the VertexCover decision problem.

The proof of Lemma 3 proceeds in Sections 4.2.1 to 4.2.6. In Section 4.2.1 we begin the Karp reduction by showing how to construct an assembly space from a vertex cover input graph. In Section 4.2.2 we build intuition around the reduction and discuss related work involving the external pointer method (EPM) [8]. In Sections 4.2.3 and 4.2.4 we show that the reduction meets the certificate requirement from Definition 1, i.e. that we have a a function $f$ for which certificates in StringAssemblyIndex are *yes* instances if and only if the corresponding VertexCover certificate is a *yes* instance. In Section 4.2.5 we show that $f$ takes $O(n^2)$ time to compute and is therefore a valid Karp reduction. Section 4.2.6 completes the proof.

[10] This is not a consequence of how we chose to define *string assembly space*. The commutativity is a fundamental attribute of assembly spaces. There is no alternative definition of a string assembly space where $[x + y \to xy]$ would imply $[y + x \to yx]$.

#### 4.2.1 Construct an Assembly Space from a Graph

Before we begin our construction of the assembly space $\Omega$, we enumerate the vertices of the graph $G = (V, E)$,

$$V = \{v_i \mid i = 1 \ldots |V|\} \tag{15}$$

We also enumerate the edges of $G$, along with their starting and terminal vertices,

$$E = \{e_j = [v_{s_j}, v_{t_j}] \mid j = 1 \ldots |E|\} \tag{16}$$

Note that a single vertex can have multiple identifications within this enumeration system. If the $j^{th}$ edge starts at the $m^{th}$ vertex and terminates at the $n^{th}$ vertex, then we could write $e_j = [v_{s_j}, v_{t_j}] = [v_m, v_n]$.

We next choose a distinct single-character string $c_i$ to correspond to each enumerated vertex $v_i \in V$. We will also define a special single-character string # that is distinct from every other single-character string. Finally, we will use $c_{s_j}$ and $c_{t_j}$ to refer to the single-character strings that correspond to the vertices of each enumerated edge $[v_{s_j}, v_{t_j}]$.

We now begin our construction of the assembly space $\Omega = (\Gamma, \phi)$ by listing all of the (string) assembly objects in $V(\Omega)$, which will be:

1. All of the single-character strings we've just defined:

$$\{\#\} \cup \{c_i \mid v_i \in V\} \tag{17}$$

2. Three composite strings for every vertex $v_i$:[11]

$$\{\#c_i,\ c_i\#,\ \#c_i\# \mid v_i \in V\} \tag{18}$$

3. One composite string for each edge $e_j = [v_{s_j}, v_{t_j}]$:

$$\{\#c_{s_j}\#c_{t_j}\# \mid e_j = [v_{s_j}, v_{t_j}] \in E\} \tag{19}$$

4. A sequence of $|V|$ composite strings, formed by starting with # and appending the $\#c_i$ strings one at a time:

$$S_0 = \# \tag{20}$$
$$S_i = (S_{i-1})(\#c_i) \qquad \forall v_i \in V \tag{21}$$

5. Another $|V|$ strings formed by further appending the $c_i\#$ strings to the final string in the previous sequence:

$$S_{|V|+1} = (S_{|V|})(c_1\#) \tag{22}$$
$$S_{|V|+i} = (S_{|V|+i-1})(c_i\#) \qquad \forall v_i \in V \tag{23}$$

6. Another $|E|$ strings formed by further appending the $\#c_{s_j}\#c_{t_j}\#$ strings:

$$S_{2|V|+1} = (S_{2|V|})(\#c_{s_1}\#c_{t_1}\#) \tag{24}$$
$$S_{2|V|+j} = (S_{2|V|+j-1})(\#c_{s_j}\#c_{t_j}\#) \quad \forall e_j \in E \tag{25}$$

The complete set of strings in $\Omega$ is thus:

$$V(\Omega) = \{\#\} \cup \{c_i, c_i\#, \#c_i, \#c_i\# \mid 1 \le i \le |V|\} \tag{26}$$
$$\cup \{\#c_{s_j}\#c_{t_j}\# \mid 1 \le j \le |E|\} \tag{27}$$
$$\cup \{S_l \mid 1 \le l \le 2|V| + |E|\} \tag{28}$$

and $\{S_l\}$ is a sequence of $2|V|+|E|+1$ strings, including, for example:

$$S_0 = \# \tag{29}$$
$$S_1 = \#\#c_i \tag{30}$$
$$S_{|V|} = \# \left(\prod_{i=1}^{|V|} \#c_i\right) \tag{31}$$
$$S_{2|V|} = \# \left(\prod_{i=1}^{|V|} \#c_i\right) \left(\prod_{i=1}^{|V|} c_i\#\right) \tag{32}$$
$$S_{2|V|+|E|} = \# \left(\prod_{i=1}^{|V|} \#c_i\right) \left(\prod_{i=1}^{|V|} c_i\#\right) \left(\prod_{j=1}^{|E|} \#c_{s_j}\#c_{t_j}\#\right) \tag{33}$$

Having thus listed the string assembly objects of $V(\Omega)$, we complete our specification of the assembly space $\Omega$ by listing is allowed assembly operations $E(\Omega)$, which will be the following:[12]

$$[c_i + \# \to c_i\#] \qquad \forall v_i \in V \tag{38}$$
$$[\# + c_i \to \#c_i] \qquad \forall v_i \in V \tag{39}$$
$$[\#c_i + \# \to \#c_i\#] \qquad \forall v_i \in V \tag{40}$$

[11] Throughout the proof of Lemma 3 and particularly in Equations (53) and (54) we use the delimiter # to organise the mapping of substrings to graph elements. This mechanism was first used by Storer and Szymanski [7, 9] to reason about the EPM of string compression. EPM compression replaces substrings in the original data with pointers to identical substrings already processed and stored in a dictionary. The substrings in the dictionary can also contain pointers to other dictionary substrings. The dictionary is analogous to an assembly space, the string to be recovered is like the target object, and the size of the "best" dictionary is reminiscent of the assembly index. The runtime complexity of EPM is $O(n \log n)$ but Storer and Szymanski proved that determining the best dictionary, i.e. one that guarantees optimal compression, is **NP-complete**. Similarly, optimal factorization (minimal phrase substitution) for Lemple-Ziv LZ77 [11] is also **NP-Complete** [7].

[12] The attentive reader may notice that Equations (38) and (39) together seem to imply that our assembly operations are non-commutative after all. This is only because we have left the commuted versions of these operations implicit. Equations (38) and (39) actually specify four dif-

$$[\#c_{s_j} + \#c_{t_j}\# \to \#c_{s_j}\#c_{t_j}\#] \qquad \forall e_j \in E \quad (41)$$
$$[\#c_{s_j}\# + c_{t_j}\# \to \#c_{s_j}\#c_{t_j}\#] \qquad \forall e_j \in E \quad (42)$$

$$[S_{i-1} + \#c_i \to S_i] \qquad \forall v_i \in V \quad (43)$$
$$[S_{|V|+i-1} + c_i\# \to S_{|V|+i}] \qquad \forall v_i \in V \quad (44)$$
$$[S_{2|V|+j-1} + \#c_{s_j}\#c_{t_j}\# \to S_{2|V|+j}] \qquad \forall e_j \in E \quad (45)$$

This completes our definition of the string assembly space $\Omega$, though it remains to be proven that $\Omega$ is indeed a string assembly space. We also define the unlabeled version of the graph $\Omega$ as $\Gamma$ and use the labels to construct an edge-labeling map $\phi$, so that $\Omega = (\Gamma, \phi)$.

To prove that $\Omega = (\Gamma, \phi)$ meets the definition of a finite assembly space, first note that the directed graph $\Gamma$ is finite and non-empty. Next, let an arbitrary edge $[x, z] \in E(\Gamma)$ be given. To exist in the directed graph $\Gamma$, this edge must correspond to one of the allowed assembly operations we specified, i.e. $[x + y \to z] \in E(\Omega)$ for some $y \in V(\Omega)$. For every such assembly operation we defined, $z$ is a string that is strictly longer than the string $x$. This means that the directed edge $[x, z]$ terminates at a string that is strictly longer than the string it began at. Because this is true of every edge in $E(\Gamma)$, there can be no cycles in $\Gamma$. We now know $\Gamma$ is a finite, non-empty, acyclic graph, so it remains only to prove that $[x + y \to z] \in E(\Omega)$ implies $[y + x \to z] \in E(\Omega)$, but that property was inherent to our method of specifying the edges and edge-labeling– each allowed assembly operation we specified implicitly specified that the commuted version was allowed as well. Thus, Definition 11 of a finite assembly space is satisfied, so $\Omega$ is indeed an assembly space.

To see that $\Omega$ is furthermore a string assembly space, note that each of the allowed assembly operations we listed has the form $[x + y \to xy]$, while the commuted version we left implicit has the form $[y + x \to xy]$ (or equivalently, $[x + y \to yx]$). In either case, Equation (10) is satisfied, and so $\Omega$ is indeed a string assembly space.

Finally, to complete the construction of our StringAssemblyIndex problem, we must define $x$ and $k'$. For reasons that will become apparent below, we choose:

$$k' = 4|V| + 2|E| + k, \qquad (46)$$

where $V$ and $E$ are from the original vertex cover graph $G = (V, E)$; and for $x$, we choose the final string in the $\{S_l\}$ sequence:

$$x = S_{2|V|+|E|} \qquad (47)$$

In the following sections, we will prove that this StringAssemblyIndex decision problem $(\Omega, x, k')$ is logically equivalent to the VertexCover decision problem $(G, k)$ from which it was constructed. We will do so by showing that a certificate for the VertexCover problem can be used to produce a certificate for the StringAssemblyIndex problem, and vice versa.

In the subsections of the proof that follow, we will refer to these problems as *the* StringAssemblyIndex *decision problem* and *the* VertexCover *decision problem*, respectively. For the duration of this proof, when we use any of the generic symbols: $G, V, E, k, \Omega, \Gamma, \phi, \{S_l\}, x, k'$, we will mean the particular values given to those symbols in this section. For example, $E$ will refer to the edges of the vertex cover graph $G = (V, E)$, and $k'$ will refer to the value $4|V| + 2|E| + k$.

#### 4.2.2 Intuition

Informally, we have designed $\Omega$ so that assembling the string $x = S_{2|V|+|E|}$ from the basic objects of $\Omega$ would require first assembling *almost* every other composite string in $V(\Omega)$. This is due to the way we defined the $\{S_l\}$ sequence, directly requiring the successive concatenation of every string in $V(\Omega)$ of the form $c_i\#$, $\#c_i$, or $\#c_{s_j}\#c_{t_j}\#$ along the path from $S_0$ to $S_{2|V|+|E|}$.

But we have not similarly made the $\#c_i\#$ strings directly required for the $\{S_l\}$ sequence. They are only indirectly required for their role in assembling the other required strings. Specifically, they are needed only to assemble the $\#c_{s_j}\#c_{t_j}\#$ strings. Critically, it will turn out that a rooted subspace of $\Omega$ is capable of assembling all of the $\#c_{s_j}\#c_{t_j}\#$ strings if and only if it contains enough of the $\#c_i\#$ strings to form a vertex cover, i.e. enough that

$$C = \{v_i \in V \mid \#c_i\# \in V(\Omega')\} \qquad (48)$$

includes at least one vertex in V incident to every edge in $E$, where $G = (V, E)$ is the graph from the VertexCover problem.[13]

ferent assembly operations:

$$[c_i + \# \to c_i\#] \qquad (34)$$
$$[\# + c_i \to c_i\#] \qquad (35)$$
$$[c_i + \# \to \#c_i] \qquad (36)$$
$$[\# + c_i \to \#c_i] \qquad (37)$$

That is to say, the string objects $c_i$ and $\#$ can be combined to assemble *either* $c_i\#$ or $\#c_i$, and their order is never relevant.

[13] This relationship will be demonstrated below. It is not meant to be clear to the reader at this time.

This bijection between vertex covers of $G$ and rooted subspaces of $\Omega$ is the core mechanism of this proof. It will allow us to use a solution to the StringAssemblyIndex problem to generate a solution to the VertexCover problem, and vice-versa, and thereby to show the problems are equivalent.

### 4.2.3 StringAssemblyIndex Certificates Map to VertexCover Certificates

Let a certificate for the StringAssemblyIndex decision problem be given, i.e. a string assembly space $\Omega'$ that is a rooted subspace of $\Omega$, contains the string $x$, and has $k'$ composite objects. We claim that the set

$$C = \{v_i \in V \mid \#c_i\# \in V(\Omega')\} \qquad (49)$$

is a vertex cover of $G = (V, E)$ with size $k$, i.e. a certificate for the VertexCover decision problem.

Our first step towards showing this claim will be to determine $B_\Omega$. This is simply the set of non-composite strings in $V(\Omega)$, those that cannot be assembled by any operation that is legal in $\Omega$. By comparing the list of strings in $V(\Omega)$ in Section 4.2.1 to the list of allowed assembly operations in $\Omega$ specified by Equations (38) to (45), we see that the only such strings are the single-character strings in $V(\Omega)$. That is:

$$B_\Omega = \{\#\} \cup \{c_i | v_i \in V\} \qquad (50)$$

Next, we will attempt to determine $V(\Omega')$, which will turn out to be all of the strings in $V(\Omega)$ minus a few of the form $\#c_i\#$. We will establish this by considering each type of string in $V(\Omega)$ in turn:

1. First, we will show that that each string in the sequence $\{S_l\}$ must be in $\Omega'$. Conceptually, this is because we've designed $\Omega$ so that the only way to assemble $x = S_{2|V|+|E|}$ is to assemble each string in the entire $S_l$ sequence. To show it rigorously, we establish the following induction:

   Suppose $S_l \in V(\Omega')$ for some $l \geq 1$. Note that $S_l \notin B_\Omega$, because it is not one of the single-character strings. Because $\Omega'$ is rooted in $\Omega$, this means that $S_l \notin B_{\Omega'}$ as well. $S_l$ is therefore a composite string in $\Omega'$, so there is an allowed operation in $E(\Omega')$ that assembles it. That operation must be one of those in Equations (43) and (45), all of which have the form $[S_{l-1} + w \to S_l]$ for some string $w \in V(\Omega)$. In order for such an operation to be allowed in $\Omega'$, is is necessary for $S_{l-1}, w \in V(\Omega')$. Thus,

   $$\forall l \geq 1, S_l \in V(\Omega') \implies S_{l-1} \in V(\Omega') \qquad (51)$$

   We began this subsection with the assumption that $x \in V(\Omega')$, i.e. $S_l \in V(\Omega')$ for $l = 2|V| + |E|$. So we can apply this induction down to $l = 1$ to conclude that $S_l \in V(\Omega')$ for all $l \geq 0$.

2. Next, consider the strings in $V(\Omega)$ of the forms $c_i\#$, $\#c_i$, and $\#c_{s_j}\#c_{t_j}\#$. Each such string is required for one of the operations by which the $S_l$ strings are assembled (Equations (43) to (45)). If any one of them were absent from $V(\Omega')$, then one of those assembly operations would not be allowed in $\Omega'$, and thus some some $S_l$ would not be a composite object in $\Omega'$. This would mean $S_l \in B_{\Omega'}$, which is false, because it is not one of the single-character strings. By this contradiction, we conclude that all of the strings in $V(\Omega)$ with these forms must also be in $V(\Omega')$.

3. Next, consider the single-character strings, $\#$ and $\{c_i\}$. If any of these were absent from $V(\Omega')$, then the operation that assembles some string of the form $c_i\#$ or $\#c_i$ would not be allowed in $\Omega'$, and one of those strings would instead be in $B_{\Omega'}$. As in the previous argument, this would yield a contradiction, so $\#$ and all of the strings $c_i$ must be in $\Omega'$.

4. Finally, consider the strings of the form $\#c_i\#$. We cannot establish exactly which of these strings are in $\Omega'$, but we *can* establish that the vertices that they correspond to comprise a vertex cover in $G$. Define

   $$C = \{v_i \in V \mid \#c_i\# \in V(\Omega')\}. \qquad (52)$$

   Now let an arbitrary edge $e_j = [v_{s_j}, v_{t_j}] \in E$ be given. We know that $\#c_{s_j}\#c_{t_j}\# \notin B_{\Omega'}$, so $\#c_{s_j}\#c_{t_j}\#$ must be a composite string in $\Omega'$. So of the two assembly operations allowed in $\Omega$ that can assemble $\#c_{s_j}\#c_{t_j}\#$, at least one must also be allowed in $\Omega'$ as well. That is, either

   $$[\#c_{s_j} + \#c_{t_j}\# \to \#c_{s_j}\#c_{t_j}\#] \in E(\Omega') \qquad (53)$$

   or

   $$[\#c_{s_j}\# + c_{t_j}\# \to \#c_{s_j}\#c_{t_j}\#] \in E(\Omega'). \qquad (54)$$

   If the first of these operations is allowed in $\Omega'$, then all of the strings involved in it must be in $V(\Omega')$. In particular, $\#c_{t_j}\# \in V(\Omega')$, which would mean $v_{t_j} \in C$. Similarly, if the second operation was allowed in $\Omega'$, it would mean $v_{s_j} \in C$. In either case, $C$ contains at least one of the vertices of the edge $e_j = [v_{s_j}, v_{t_j}]$.

   Because this is true for every edge $e_j \in E$, $C$ is a vertex cover of $G$.

We will further prove that $|C| = k$. To do so, first note that the steps above have determined which strings are in $\Omega'$:

$$V(\Omega') = \{S_l \mid l = 1 \ldots 2|V| + |E|\} \tag{55}$$
$$\cup \{c_i\#, \#c_i \mid i = 1 \ldots |V|\} \tag{56}$$
$$\cup \{\#c_{s_j}\#c_{t_j}\# \mid j = 1 \ldots |E|\} \tag{57}$$
$$\cup \{\#\} \cup \{c_i \mid i = 1 \ldots |V|\} \tag{58}$$
$$\cup \{\#c_i\# \mid v_i \in C\} \tag{59}$$

We can therefore calculate

$$|V(\Omega')| = 1 + 5|V| + 2|E| + |C| \tag{60}$$

We've established above that $B_\Omega$ consists all of the single character strings in $V(\Omega)$, and that all of those strings are in $V(\Omega')$ as well. Each such single-character string is a basic object in $\Omega'$, and no other objects can be basic objects in $\Omega'$ because $\Omega'$ is rooted in $\Omega$. Therefore,

$$B_{\Omega'} = \{\#\} \cup \{c_i | v_i \in V\} \tag{61}$$
$$\therefore |B_{\Omega'}| = 1 + |V| \tag{62}$$

Putting these together yields

$$|V(\Omega')| - |B_{\Omega'}| = 4|V| + 2|E| + |C| \tag{63}$$

Because we began this section with the assumption that $\Omega'$ has $k'$ composite objects, we have

$$k' = 4|V| + 2|E| + |C| \tag{64}$$
$$\therefore 4|V| + 2|E| + k = 4|V| + 2|E| + |C| \tag{65}$$
$$\therefore k = |C| \tag{66}$$

Thus, $C$ is a vertex cover of $G$ with size $k$, as desired. It is a certificate to the original VERTEXCOVER decision problem.

#### 4.2.4 VERTEXCOVER Certificates Map to STRINGASSEMBLYINDEX Certificates

We will now show the other direction. Let a certificate for the VERTEXCOVER decision problem be given, i.e. a vertex cover $C$ of the graph $G = (V, E)$ with $|C| = k$. We will use $C$ to construct a string assembly space $\Omega' = (\Gamma', \phi')$ that is a certificate for the STRINGASSEMBLYINDEX decision problem.

Conceptually, the subspace $\Omega' = (\Gamma', \phi')$ we will construct is just $\Omega = (\Gamma, \phi)$ with some of the $\#c_i\#$ strings removed, along with the assembly operations that involve those strings. Formally, we define the vertices of the graph $\Gamma'$ to be

$$V(\Gamma') = V(\Gamma) \setminus \{\#c_i\# | v_i \in V - C\} \tag{67}$$

For the edges of $\Gamma'$, we simply include every edge from $\Gamma$ that we can. That is, we include every edge from $\Gamma$ whose vertices and label are all included in $\Gamma'$:

$$E(\Gamma') = \{e = [s, t] \mid e \in E(\Gamma) \text{ and } s, t, \phi(e) \in V(\Gamma')\} \tag{68}$$

Finally, we simply define the edge-labeling map $\phi'$ to be consistent with $\phi$:

$$\phi'(e) = \phi(e) \;\; \forall\, e \in E(\Gamma') \tag{69}$$

Note that this construction implies that $\Omega'$ has all the same allowed operations as $\Omega$, unless it is missing one of the strings involved in one of those operations. That is,

$$[x + y \to z] \in E(\Omega') \tag{70}$$
$$\iff [x + y \to z] \in E(\Omega) \wedge x, y, z \in V(\Omega'). \tag{71}$$

Now that we have defined $\Omega' = (\Gamma', \phi')$, we need to prove that it really is an assembly space. We know the graph $\Gamma'$ is finite and acyclic because it is a subgraph of $\Gamma$, and we know it is non-empty because it contains at least $\#$. Now suppose that $[x + y \to z] \in E(\Omega')$. Then Equation (71) tells us that $x, y, z \in V(\Omega')$ and $[x + y \to z] \in E(\Omega)$. Because $\Omega$ is an assembly space, the commuted version of this operation is also allowed, i.e. $[y + x \to z] \in E(\Omega)$. Now applying Equation (71) again tells us that $[y + x \to z] \in E(\Omega')$ as well, so $\Omega'$ meets the definition of a finite assembly space. To show that it is specifically a string assembly space, note that all of its assembly objects are strings, and that $E(\Omega') \subset E(\Omega)$. Because the assembly operations in $E(\Omega)$ obey Equation (10), those in $E(\Omega')$ do as well, and so Definition 19 is satisfied.

Because the graph $\Gamma'$ is a subgraph of $\Gamma$ and the edge-labeling functions $\phi$ and $\phi'$ are consistent, $\Omega'$ is furthermore an assembly subspace of $\Omega$. To show that $\Omega'$ is rooted in $\Omega$, we will show that $B_{\Omega'} = B_\Omega$. In the previous section, we found the basis of $\Omega$ to be all of its single-character strings. Each of these strings is in $\Omega'$ as well by Equation (67), and each is reachable only from itself in $\Omega'$. Each single-character string is therefore a member of $B_{\Omega'}$.

To show that none of the other strings in $\Omega'$ can be in $B_{\Omega'}$, we will consider each in turn:

- Consider the string $c_i\#$ for any $1 \leq i \leq |V|$. By Equation (67), $c_i, \#, c_i\# \in V(\Omega')$, and Equation (38) says $[c_i + \# \to c_i\#] \in E(\Omega)$, so by Equation (71), $[c_i + \# \to c_i\#] \in E(\Omega')$. Since $c_i\#$ is thus a composite object in $\Omega'$, $c_i\# \notin B_{\Omega'}$. A symmetric argument tells us that $\#c_i \notin B_{\Omega'}$.

- Next, consider the string $\#c_i\#$ for any $1 \leq i \leq |V|$. If $c_i \notin C$, then by Equation (67), $\#c_i\#$ is not in $V(\Omega')$, and so it cannot be in $B_{\Omega'}$. Alternatively, if $v_i \in C$, then $\#c_i\#$ is in $V(\Omega')$, so because $\#c_i$ and $\#$ are also in $V(\Omega')$, the assembly operation

$$[\#c_i + \# \rightarrow \#c_i\#]$$

that was allowed in $\Omega$ (Equation (40)) is also allowed in $\Omega'$. Again, $\#c_i\#$ is reachable from another string in $\Omega'$, so it is not in $B_{\Omega'}$.

- Next, consider the string $\#c_{s_j}\#c_{t_j}\#$ for any $1 \leq j \leq |E|$. Because $C$ is a vertex cover of the graph $G = (V, E)$ and $[v_{s_j}, v_{t_j}]$ is one of its edges, $C$ must contain at least one of its vertices. If $v_{s_j} \in C$, then $\#c_{s_j}\# \in V(\Omega')$ by Equation (67). Because $c_{t_j}\#, \#c_{s_j}\#c_{t_j}\# \in V(\Omega')$ as well, the assembly operation

$$[\#c_{s_j}\# + c_{t_j}\# \rightarrow \#c_{s_j}\#c_{t_j}\#] \tag{72}$$

that was allowed in $\Omega$ (Equation (42)) is also allowed in $\Omega'$, so $\#c_{s_j}\#c_{t_j}\# \notin B_{\Omega'}$ by the same argument as the previous bullet point. If instead $v_{t_j} \in C$, we follow the same logic to conclude that the assembly operation

$$[\#c_{s_j} + \#c_{t_j}\# \rightarrow \#c_{s_j}\#c_{t_j}\#] \tag{73}$$

from Equation (41) is allowed in $\Omega'$, and we again conclude that $\#c_{s_j}\#c_{t_j}\# \notin B_{\Omega'}$.

- Finally, consider the string $S_l$ for any $1 \leq l \leq |V|$. By the same argument as in the previous bullet points, the assembly operation

$$[S_{l-1} + \#c_l \rightarrow S_l] \tag{74}$$

from Equation (43) must be allowed in $\Omega'$, so $S_l$ is reachable from $S_{l-1}$ in $\Omega'$, and therefore is not in $B_{\Omega'}$. If $|V| < l \leq 2|V| + |E|$, the argument is identical, but uses either Equation (44) or Equation (45).

That's all the strings in $\Omega$, and hence all the strings that can be in the basis of $\Omega'$. We conclude that $B_{\Omega'}$ consists of only the single-character strings,

$$B_{\Omega'} = \{\#\} \cup \{c_i | i = 1 \ldots |V|\} = B_\Omega$$

We now know that $\Omega'$ is a rooted assembly subspace of $\Omega$ that contains $x$. All that remains is to calculate its number of composite objects. We can tally up the total number of strings in $V(\Omega)$ using the list on page 6:

$$|V(\Omega)| = 1 + 6|V| + 2|E| \tag{75}$$

so together with Equation (67), we get

$$\begin{aligned} |V(\Omega')| &= |V(\Omega)| - (|V| - |C|) && (76)\\ &= (1 + 6|V| + 2|E|) - (|V| - |C|) && (77)\\ &= 1 + 5|V| + 2|E| + |C| && (78) \end{aligned}$$

$$\begin{aligned} \therefore |V(\Omega')| &- |B_{\Omega'}| && (79)\\ &= (1 + 5|V| + 2|E| + |C|) - (1 + |V|) && (80)\\ &= 4|V| + 2|E| + |C| && (81) \end{aligned}$$

and we began this section with the assumption that $|C| = k$, so we conclude that the number of composite objects in $\Omega'$ is

$$|V(\Omega')| - |B_{\Omega'}| = 4|V| + 2|E| + k = k' \tag{82}$$

Putting this all together, $\Omega'$ is a string assembly space that is rooted in $\Omega$, contains $x$, and has $k'$ composite objects. That is, it is a certificate to the StringAssemblyIndex decision problem, as desired.

#### 4.2.5 Calculating the Runtime

By combining the last two subsections, we know that our two decision problems always have the same answer; a certificate for one exists if and only if a certificate for the other exists. We have thus shown a way to reduce an arbitrary VertexCover decision problem to an equivalent StringAssemblyIndex decision problem.

This reduction requires only $O(N^2)$ time, where $N = |V| + |E|$ is the input size of the VertexCover problem. We only need to read the specification of the graph $G$ as input and produce the specification of the assembly space $\Omega = (\Gamma, \phi)$, along with $x$ and $k'$. This can be done in $O(N^2)$ time by following the steps beginning on page 6 to list the string assembly objects in $\Omega$, and then following Equations (38) and (45) to list its allowed assembly operations. The list of allowed assembly operations already constitutes an unambiguous implicit specification of the edges of $\Omega$ and its edge-labeling map $\phi$, but each assembly operation can also be transformed in an explicit specification of the two corresponding edges and their labels in $O(1)$ time.

The number of strings in $\Omega$ is $1 + 6|V| + 2|E|$, and the number of allowed assembly operations is $5|V| + 2|E|$. The reduction process requires no additional computation that scales with the input size; it is only copying strings, and each string or assembly operation can be written in $O(1)$ time. So there are $O(|V| + |E|) = O(N)$ statements to write, each of length $O(N)$. So reducing the VertexCover decision problem to our StringAssemblyIndex decision problem requires only $O(N^2)$ time.

#### 4.2.6 StringAssemblyIndex is an NP-Hard Problem

Because an arbitrary VertexCover decision problem can be reduced to a StringAssemblyIndex decision problem in polynomial-time (Definition 1), StringAssemblyIndex is **NP-Hard**. □

## 5 AssemblyIndex is NP-Complete

Completing the proof that both AssemblyIndex problems are **NP-Complete** is now only a formality. The remaining steps follow immediately from the fact that StringAssemblyIndex problems are a specific subclass of AssemblyIndex problems and that each allow their optimisation versions to be solved with only a polynomial number calls to the decision versions.

**Corollary 1** (StringAssemblyIndex is in **NP**). *A proposed solution (certificate) to a* StringAssemblyIndex *decision problem can be checked in polynomial time.*

*Proof.* A certificate for a StringAssemblyIndex decision problem is also a certificate for a (general) AssemblyIndex decision problem. Lemma 2 tells us that any certificate in the latter class can be checked in polynomial time, so the result follows. □

**Corollary 2** (AssemblyIndex is **NP-Hard**). *Any decision problem in* **NP** *can be reduced to a (general)* AssemblyIndex *decision problem in polynomial time.*

*Proof.* By Lemma 3, any decision problem in **NP** can be reduced to a StringAssemblyIndex decision problem in polynomial time. Any such problem is also a (general) AssemblyIndex decision problem, so the result follows. □

**Theorem 1** (AssemblyIndex is **NP-Complete**). *The* AssemblyIndex *problem is* **NP-Complete**.

*Proof.* With Definition 2, Lemma 2, and Corollary 2 the result follows. □

**Theorem 2** (StringAssemblyIndex is **NP-Complete**). *The* StringAssemblyIndex *problem is* **NP-Complete**.

*Proof.* Combining Definition 2, Lemma 3, and Corollary 1 completes the proof. □

## Acknowledgements

We thank Ulysse Prieto for valuable discussions of string pathway complexity and its relation to EPM.